\documentclass[12pt,preprint]{aastex}

\usepackage{rotating}

\slugcomment{v0 initial draft}

\def\sample{83 }

\shorttitle{Disk opacity from occulting pairs}
\shortauthors{Holwerda et al.}

\begin{document}

\title{Spiral Disk Opacity from Occulting Galaxy Pairs in the Sloan Digital Sky Survey}
\author{B. W. Holwerda\altaffilmark{1}, W. C. Keel\altaffilmark{2} \and A. Bolton\altaffilmark{3}}
\email{holwerda@stsci.edu}

\altaffiltext{1}{Space Telescope Science Institute, 3700 San Martin Drive, Baltimore, MD 21218, USA}
\altaffiltext{2}{Department of Physics and Astronomy, University of Alabama, Box 870324, University of Alabama, Tuscaloosa, AL 35487-0324, USA}
\altaffiltext{3}{Harvard-Smithsonian Center for Astrophysics, 60 Garden St. MS-20, Cambridge, MA 02138, USA}

\begin{abstract}
A spiral galaxy partially ovelapping a more distant elliptical offers an unique opportunity 
to measure the dust extinction in the foreground spiral. From the Sloan Digital Sky Survey 
DR4 spectroscopic sample, we selected \sample occulting galaxy pairs and measured 
disk opacity over the redshift range z = 0.0-0.2 with the goal to determine the recent evolution 
of disk dust opacity. 
The enrichment of the ISM changes over the lifetime of a disk and it is reasonable to expect 
the dust extinction properties of spiral disks as a whole to change over their lifetime. When 
they do, the change will affect our measurements of galaxies over the observable universe. 

From the SDSS pairs we conclude that spiral disks show evidence of extinction to $\sim$ 2 
effective radii. However, no evidence for recent evolution of disk opacity is evident, due to 
the limited redshift range and our inability to distinguish other factors on disk opacity such 
as the presence of spiral arms and Hubble type. Such effects also mask any relation between 
surface brightness and optical depth that has been found in nearby galaxies.
Hence, we conclude that the SDSS spectral catalog is an excellent way to find occulting pairs 
and construct a uniform local sample. However, higher resolution than the SDSS images is 
needed to disentangle the effects of spiral arms and Hubble type from evolution since z = 0.2.

\end{abstract}


\section{\label{s:intro}Introduction}

The role of extinction may well change across the now-observable range of
redshifts, as heavy elements are injected into the ISM but the
reservoir of gas within a galaxy is progressively cycled through 
stars. Thus the history of dust within galaxies is intimately
linked to the history of the overall star-formation rate, which
has been at least partly specified through observations of several
deep fields with HST and other instruments. \cite{Calzetti99}
incorporated both SFR and abundance constraints to examine the possible
histories of galaxy extinction, finding that these constraints allow
two classes of solutions. These have markedly different redshift behavior,
one with a peak extinction early on ($z \approx 3$), and the other 
building up dust more gradually with a peak extinction as late as $z=1$.

The typical dust mass in distant galaxies is very much a function of the selected 
sample. Far-infrared selected samples point to more Arp-220 like dust prominence 
\citep{RR05}, optical/UV selected samples point to disks very similar to the local 
ones \citep{Sajina06} and Lyman-$\alpha$ galaxies point to low-extinction disks 
\citep{Nilsson07}.

The observational situation at large redshifts is still ill-constrained, so far based on
(i) modeling the spectral energy distributions (SEDs) of galaxies at high $z$ \citep{RR03,RR05,Sajina06,Nilsson07},
(ii) measuring differences in the SEDs of lensed quasars in which one image
is formed deep within the lens system \citep[e.g, ][]{Nadeau91,Falco99, Motta02, Eliasdottir06}, 
(iii) correlating the colors of quasars with metal-line absorption systems 
compared to average QSO SEDs \citep{Hopkins04}, and 
(iv) the extinction fits to Sn1a lightcurves \citep[e.g., ][]{Knop03,Jha07}.
To date, the SED models have yielded the most results: \cite{RR03} find 
a peak extinction at z $\sim$ 1 and lower beyond that --no extinction at z = 6-7 
according to \cite{Yan05}. Yet \cite{Vijh03} find a strong correction for the SFR in the 
early universe due to dust attenuation by an LMC-type dust. The QSO SED's 
match the SMC's extinction law with most of the extinction in the nucleus itself. 
Most of the lensing systems are early-types however.
Lensing measurements find a wide range in the value of $R_V$ and the 
SN1a measurements find no change in reddening.

The first three methods are very vulnerable to color and surface brightness selection 
effects and the SED models are very dependent on the assumed geometry of 
ISM and stars. Partially overlapping galaxies offer a alternative and more direct way to approach the
question of the evolution of dusty ISM in spirals.

\section{\label{s:method}The occulting galaxy pairs technique}

\begin{figure}
\centering
\plotone{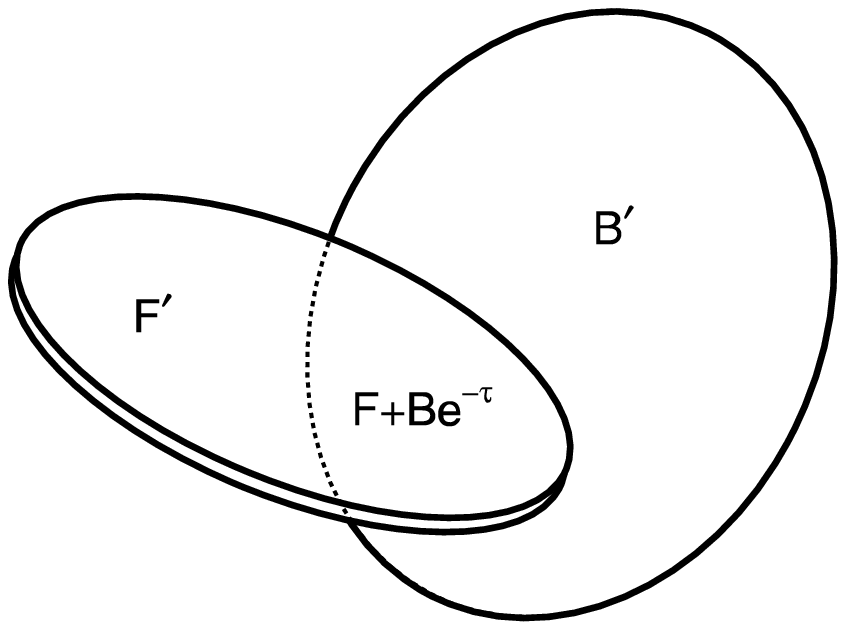}
\caption{\label{f:method} A schematic of the ideal occulting pair. The background galaxy is a symmetric elliptical galaxy and the foreground galaxy a symmetric spiral. In the overlap region, the elliptical is bright enough to provide an extinction signal but the elliptical is not completely hidden behind the spiral so the contribution of the elliptical to the overlap region can be estimated from the non-occulted part.}
\end{figure}

An occultinig galaxy pair --a foreground spiral partially covering a background galaxy-- 
can be used up to high redshift to determine the opacity of the spiral. The flux 
contributions in the overlap region from both galaxies are estimated from the 
non-overlap parts (Figure \ref{f:method}).
From a single image, we obtain the non-overlapping flux from the foreground 
spiral (F'), the non-overlap background flux (B') and the flux from the overlap 
region ($F+Be^{-\tau}$). We can now estimate the optical depth ($\tau'$) from 
these three observables: $e^{-\tau'} = ( [F+Be^{-\tau}] - F')  /  B'$. 

As the background galaxy, a partially occulted ellipical is ideal because its light 
profile is very symmetric: the assumption that the non-occulted part is a good 
approximation of the occulted part is a reasonable one. This leaves the 
assumption that the foreground spiral is symmetric as the predominant source 
of uncertainty. The method has been extensively used in the local universe which 
gives us a direct comparison set for more distant measurements. 

The occulting galaxy technique was originally proposed by \cite{kw92} to probe 
spiral disk extinction in the local universe, and in the following decade the known 
nearby overlapping-galaxy pairs were exhausted, using both ground-based imaging 
\citep{Andredakis92, Berlind97, kw99a, kw00a}, spectroscopy \citep{kw00b} 
and HST/WFPC2 imaging \citep{kw01a, kw01b, Elmegreen01}. The ideal pair, 
described by \cite{kw92}, is a bright elliptical partially behind a symmetric, face-on 
spiral. Since only a few pairs were known in the local universe, these studies 
used a great variety of background galaxies. Their results included radial extinction 
plots for the extinction across the entire height of the disk. ? 
The results follows a gray extinction law when taken over large 
regions but approaches a typical Galactic extinction law for scales smaller than 
100 pc \citep{kw01a}. Some indication that the cloud sizes are fractal in nature was found by 
\cite{kw01a}. \cite{Holwerda05b} used counts of distant galaxies seen in
HST images of nearby spirals to independently confirm the values for disk extinction from 
occulting pairs.

From the SDSS DR4 spectroscopic sample we have selected \sample occulting pairs 
with an elliptical as the background galaxy --the ideal configuration. The foreground 
spiral galaxies span a range in redshift (z=0.01-0.3). Starting with spectroscopic 
identification of both sources and their redshifts represents a significant improvement 
in reliability and quality of the pairs. The ideal occulting pair --spiral over elliptical-- 
provides the best estimate of dust content of a the spiral disk (Figure \ref{f:method}).
An additional advantage of this SDSS sample is a high separation in redshift 
($\Delta z$) of background and foreground galaxy; there are fewer issues with 
scatter of background light and interacting pairs.

This SDSS sample has selection properties more like 
what we are likely to get in the high-z fields than our carefully culled 
previous overlap sample (carefully culled to reduce the incidence of 
galaxies with asymmetry issues, for example) and this, while the bulk 
measurements will have larger errors, is perhaps a better starting point if we
want to look for evolution of the disk extinction. Also, this technique,
while very likely suboptimal for some individual marginal cases where
an analysis ``by hand'' could do better, is more like what one can try
on the large samples of poorly-resolved systems at higher redshift 
($z>0.3$).

\section{\label{s:sample}Sample selection}

\cite{Warren96,Bolton04} describe a technique to select high-redshift spiral galaxies 
lensed by a foreground elliptical galaxy from SDSS DR4 spectroscopic 
sample for HST snapshot follow-up \citep{Koopmanssnap,Boltonsnap}. 
They select red, absorption-dominated spectra --typical for ellipticals-- that also display 
multiple high-redshift emission lines associated with the background lensed 
spiral galaxy \citep{Bolton06, Treu06, Koopmans06, Gavazzi07}. 
It is a straightforward matter to implement the reverse of this algorithm 
to select absorption-dominated spectra with emission lines at {\it lower} redshifts, 
to find spiral disks with a bright background elliptical behind them.

Objects were selected from the fourth datarelease of the Sloan Digital Sky Survey
spectra if they met the following criteria:
\vspace{-5pt}
\begin{itemize}
\itemsep=0.2pt
\item[1.] The redshift is successfully found by the Princeton 1D spectral pipeline \\ 
(http://spectro.princeton.edu/).
\item[2.] The objects is classified spectroscopically as a galaxy by this pipeline:
the $\chi^2$ fit is better for a galaxy spectrum than a star or QSO.
\item[3.] Rest-frame H$\alpha$ equivalent width less than 4 \AA ~ to select ellipticals.
\item[4.] Redshift less than 0.4 to enable the H$\alpha$ equivalent width cut in the optical band.
\item[5.] At least 3 of 5 lines ([OII], H$\beta$, [OIII], [OIII] and H$\alpha$) 
are detected at a redshift greater than 0.01 but less than the elliptical's redshift.
\end{itemize}
\vspace{-5pt}

\begin{figure}
\centering
\plotone{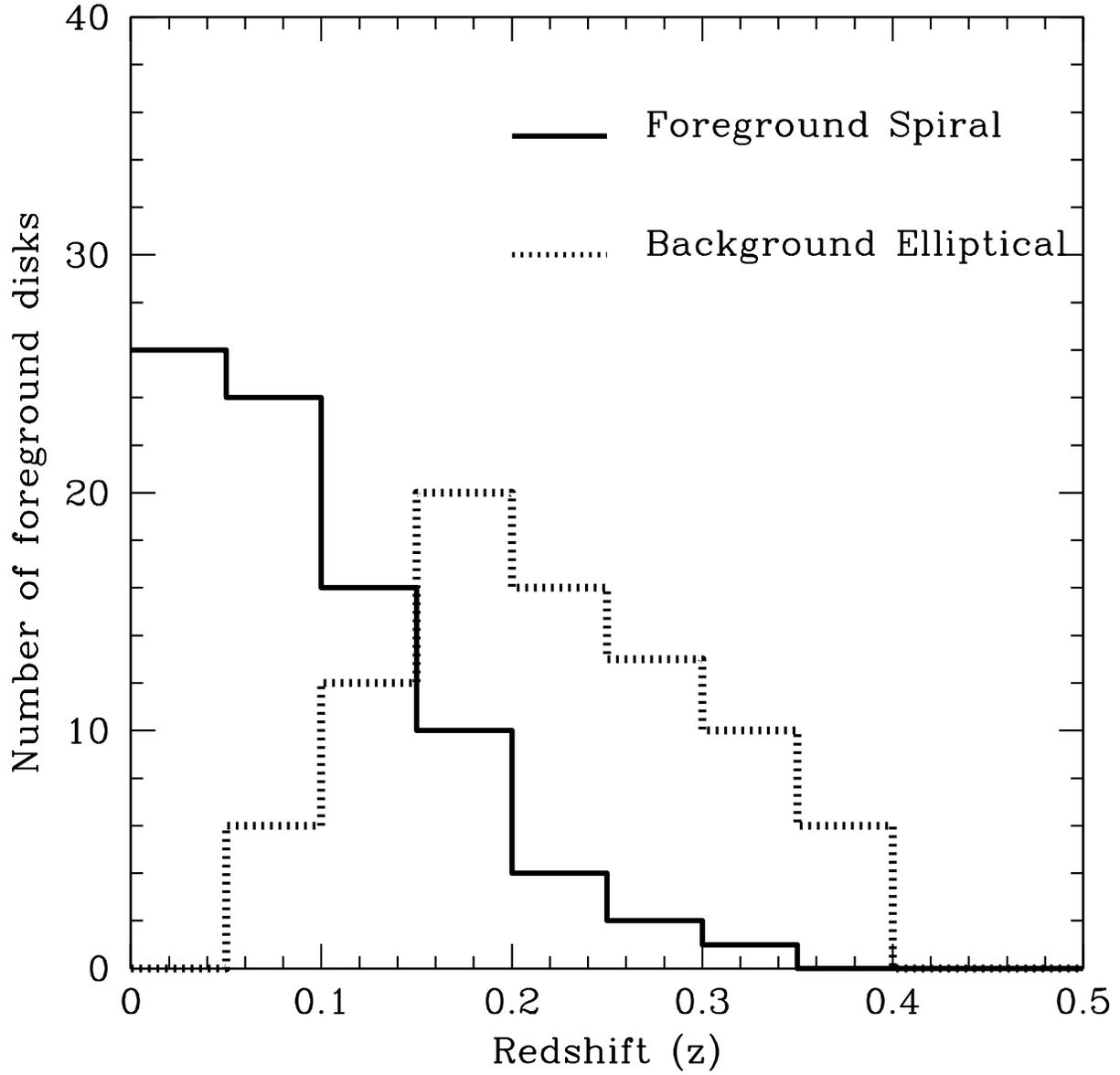}
\caption{\label{f:z}The histogram of redshift distribution of both the background and foreground galaxy. The majority of pairs is nearby (z$<$0.1). The spectroscopic selection from the SDSS limits the pairs to closer than z = 0.4. }
\end{figure}

The reversed algorithm yielded 118 candidate occulting pairs in the SDSS DR4 spectral 
sample. Both galaxies are within the 3'' aperture of the SDSS fiber, ensuring a small 
angular separation of the occulting pair members. In addition, we have visually verified 
the suitability of each pair and we picked only those that have the ideal pair geometry 
(Figure \ref{f:method}). The resulting sample is \sample ideal pairs. 
The Sloan spectroscopy limits us effectively to galaxies with redshifts less than 
0.4 ($z<0.4$), and most of the spirals are closer than 0.3 ($z< 0.3$).
Figure \ref{f:z} shows the distribution of redshifts for the foreground and 
background galaxies. The majority of the foreground galaxies is nearby (z$<$0.1) 
The SDSS DR4 spectral catalog information is listed in Table \ref{t:sample}.
These spirals, backlit by a bright symmetric elliptical, enable us to measure 
the extinction to the highest degree of accuracy possible with this technique.

The Sloan Digital Sky Survey \citep[SDSS]{York00} has mapped one-quarter 
of the entire sky mainly around the north galactic cap in five bands, {\it u', g', r', i', z'} 
\cite{Smith02, Fukugita96}. SDSS imaging is obtained using a drift-scanning 
mosaic CCD camera \citep{Gunn98} with a pixel size of $0\farcs396$. 
We obtained all five bands of the night sky (``corrected frames'') in fits format using 
the SDSS SkyServer DR5 (http://cas.sdss.org/dr5/en/tools/chart/chart.asp). 
The corrected frames, having been bias subtracted, flat-fielded, and purged of bright 
stars are stored at SDSS in integer format to save disk space. The pixel values 
get randomised appropriately before being rounded to make sure that the statistics 
of the background counts are what they should be. An additional offset (SOFTBIAS) 
of 1000 counts is added to each pixel to avoid negative pixel values and should be 
subtracted together with the sky value. We used only single SDSS scans for our analysis 
(no pair was on the dividing line between scans).

\section{\label{s:fit}Fit to the occulting pair}

The uniform approach to the SDSS images of these pairs is to fit the images with the 
central x and y position of the foreground and background galaxy as free parameters as well 
as the angle of rotation of both galaxies, 6 free parameters total ($\rm x_{fg}, y_{fg}, pa_{fg},  
x_{bg}, y_{bg}, pa_{bg}$). The best fit criteria is a minimal residual image, the original image 
with the two rotated galaxies subtracted. We used a new fit for each SDSS filter. Because the 
image is used to model itself, we do not need to take the SDSS point spread function into account.

To identify objects in the field and the members of the pair, we ran source extractor v2.5 
\citep{SE}\footnote{See also the user manual, \cite{seman}.}. 
If both members of the pair are separated  and identified in the segmentation by source 
extractor, we can proceed with the fit. We use the assignments of pixels by source extractor 
to mask non-pair object. A 100 by 100 pixel postage stamp is cut from the SDSS scan 
around the pair and used for the further fit with the IDL routine {\it mpfit2dfun}. 
 The mask for each pair member is padded by smoothing it with a 3 pixel 
wide boxcar. Typical galaxy sizes are several hundred pixels. We use the padded 
masks to determine which pixels are to be used for the 
fit. The model subtracted from the original image is the foreground galaxy rotated, plus 
the background galaxy rotated with the sky value subtracted from the original image. 
The source extractor catalog used to make the segmentation image also provides the 
effective radius, ellipticity and flux for both the foreground galaxy and background galaxy. 
The ellipticity is converted into an inclination estimate ($ i ~ = ~ atan(B/A)$).

We apply two apertures to the extinction map based on the fits, each 5$\times$5 pixels. 
In the g-band images, we visually identified a good position for an aperture. An additional aperture 
position can be identified in the extinction map by using the point of maximum extinction as the 
center of the second aperture. 
Extinction values for the visual aperture are presented in Tables \ref{t:tau} for the SDSS filters. 
The disk opacities have been corrected for the inclination ($\times cos(i)$).

The inclusion of an automatically selected aperture position was to test whether or not this method 
can be easily automated to apply on large samples of near pairs. It appears that a visual 
placement of the aperture remains optimal.

The fit fails in case source extractor did not segment the pair into two separate objects or when 
there is no signal in the visual aperture. Negative extinction values occur when the galaxies 
are significantly asymmetric but the fit does not explicitly fail. We include these negative opacity 
values because it gives an indication of the dominant remaining uncertainty in these measurements. 

\begin{figure}
\centering
\plotone{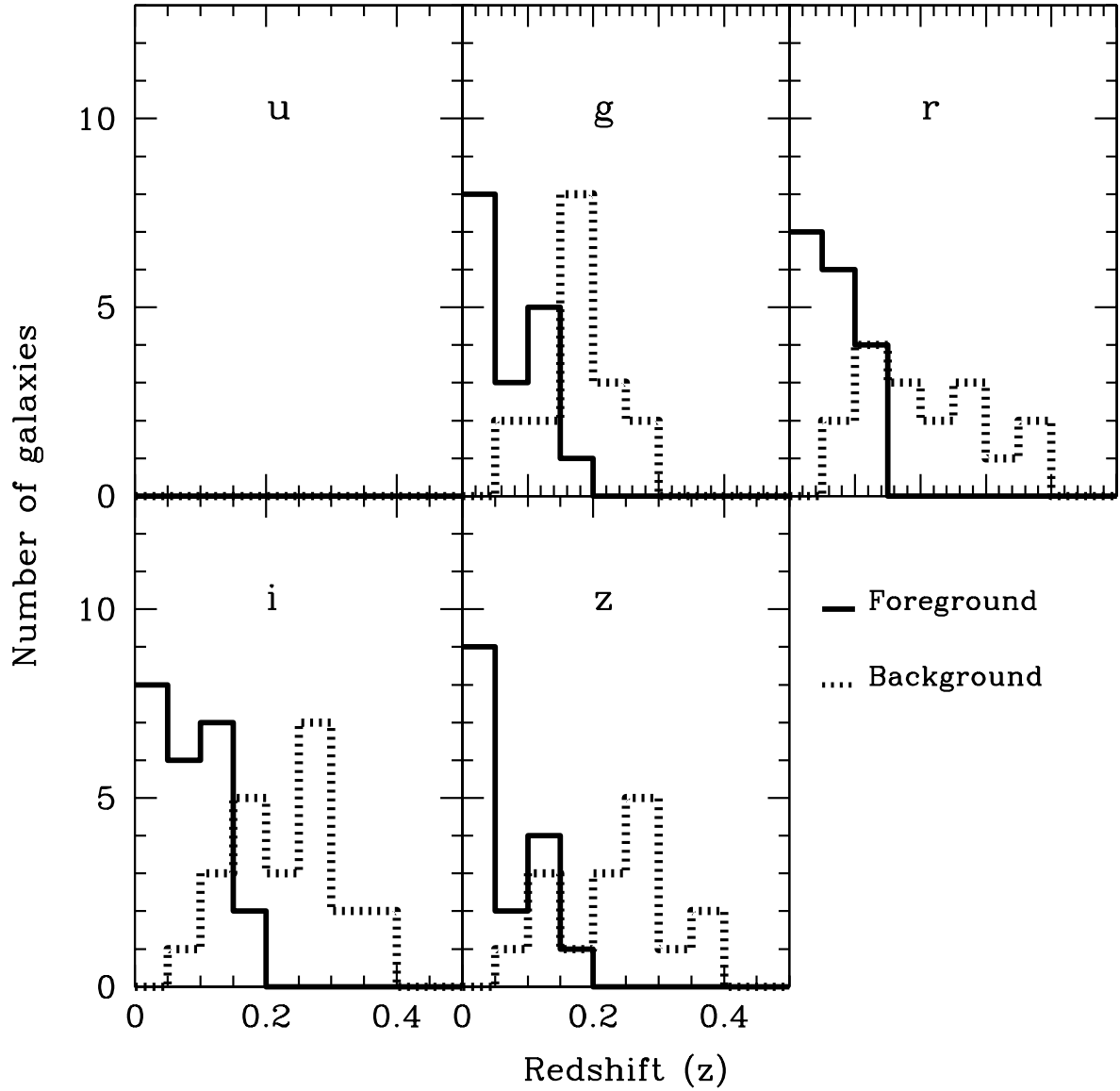}
\caption{\label{f:zfit}The histogram of redshift distribution of both the background and foreground galaxies in pairs for which a successful fit is found. No good fits are obtained for the u band and only disks closer than z=0.2 could be fitted.}
\end{figure}

Figure \ref{f:zfit} shows the redshift distributions for pairs with a successful fit. 
An opacity measurement of a spiral disk from SDSS images is feasible for disks closer than z=0.2. 
The fit fails for more distant disks as the pair cannot be resolved into separate objects.

\section{\label{s:res}Results}

The optical depth value of a spiral disks depends a great many factors, radius, arm presence, Hubble 
type and possibly disk luminosity. In this section we explore some relations between disk opacities and 
other disk characteristics, radius, disk luminosity and surface brightness. Finally, we present radial 
plots based on stacked SDSS filters together to improve signal-to-noise.

\subsection{\label{ss:ra}Radial profiles}

\begin{figure}
\centering
\plotone{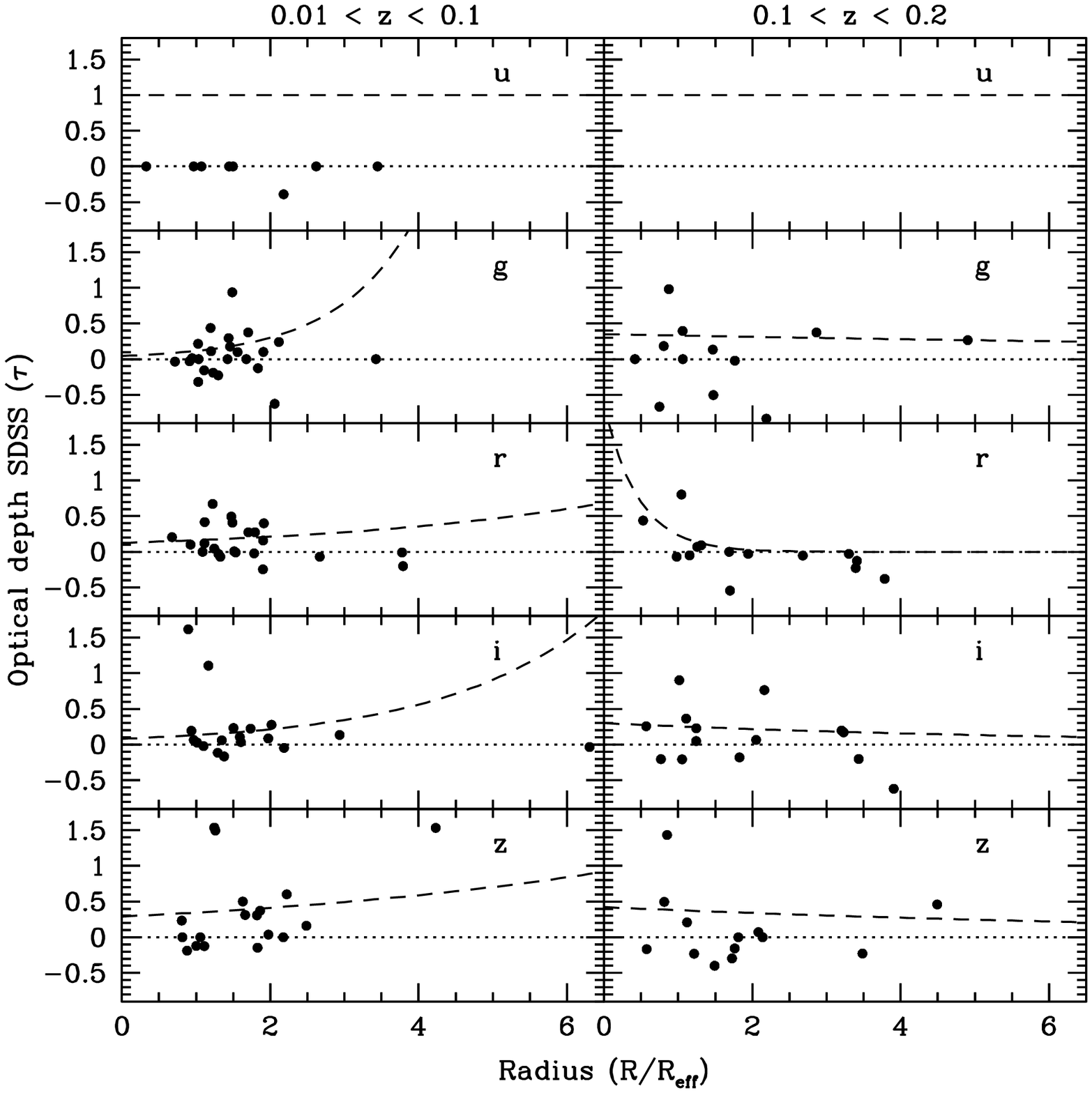}
\caption{\label{f:ra}Disk opacity in the visual aperture as a function of radius in all five SDSS filters; {\bf top:} z$<$0.05,  {\bf middle:} 0.05$>$z$>$0.1,  {\bf bottom:} z$>$0.1. The dashed lines represent and exponential fit to the positive optically thin opacity values. }
\end{figure}

\begin{figure}
\centering
\plotone{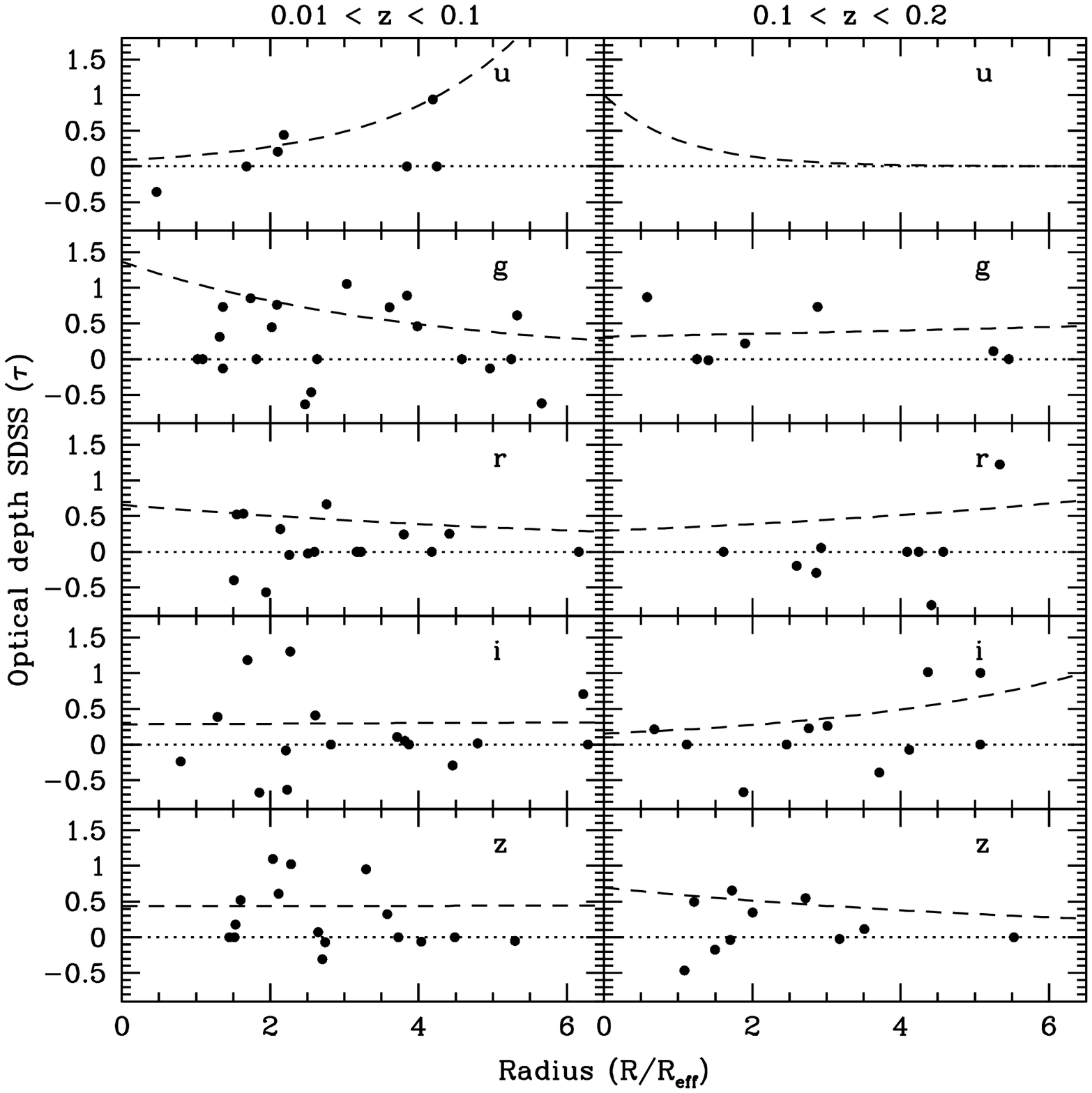}
\caption{\label{f:ra:opt}Disk opacity in the automatic aperture as a function of radius in all five SDSS filters; {\bf top:} z$<$0.05,  {\bf middle:} 0.05$>$z$>$0.1,  {\bf bottom:} z$>$0.1. The dashed lines represent and exponential fit to the positive optically thin opacity values. }
\end{figure}

Figure \ref{f:ra} shows the radial profiles based on the visual apertures for all five SDSS filters 
(g', r', i', u', z') and split up into three redshift ranges ($z<0.05$, $0.05<z<0.1$ and $z>0.1$). 
Figure \ref{f:ra:opt} shows the same plot for the automated aperture. The increase of scatter in 
the automatic aperture does indicate that the visual aperture is better placed to measure disk opacity.
The opacity values have been corrected for inclination ($\times cos(i)$). The r' and i' filter has 
the best extinction signal. Notably, the u' band fits are all failures as the early type 
background galaxy is too faint in this band and the foreground galaxy possibly too irregular. 
The reason that the most extinction signal is in the r' and i' filters is because these are blue enough 
to exhibit strong dust extinction but red enough to benefit from a symmetric distribution of the 
light in both galaxies. Bluer filters are more asymmetric  due to localized star formation in the 
foreground galaxy's spiral arms. This is in agreement with HST/WFPC2 results in local galaxies \citep{kw01a,kw01b}.

The resolution of the SDSS for these pairs is unfortunately not sufficient to distinguish between 
arm and disk sections similar to previous studies in the very local universe. In order to obtain 
similar quality measurements, one would greater resolving power such as HST.
The opacity values in Figure \ref{f:ra} are therefore a mix between the arm and disk values.

Similarly, distinguishing Hubble subtypes for these galaxies is equally impossible and hence the 
opacities represent also mix between earlier and later type spiral, which most likely have different 
opacities \citep{kw00a,Holwerda05b}. 

Because of the many other influences and the small span of the redshift range in Figure \ref{f:ra} it is 
impossible to conclude if there is any evolution in the opacity of spiral disks from the single-filter SDSS data. 
However, we can conclude that optical disks show evidence of extinction up to twice their effective 
radius in all SDSS filters except u, which is due to low S/N from the red background galaxy.

\subsection{\label{ss:fit}Exponential fits}

In both Figures \ref{f:ra} and \ref{f:ra:opt} we show the least-square fit of an exponential disk to the positive
points (fit-parametes are listed in Table \ref{t:vis_exp} and \ref{t:opt_exp}). The fits are poor and unrealistic 
(increasing with radius) for many of the SDSS filters. For a better fit, one would need better S/N and 
distinguish between arm and disk sections .

\subsection{\label{ss:mt}Disk opacity and luminosity}

\begin{figure}
\centering
\plotone{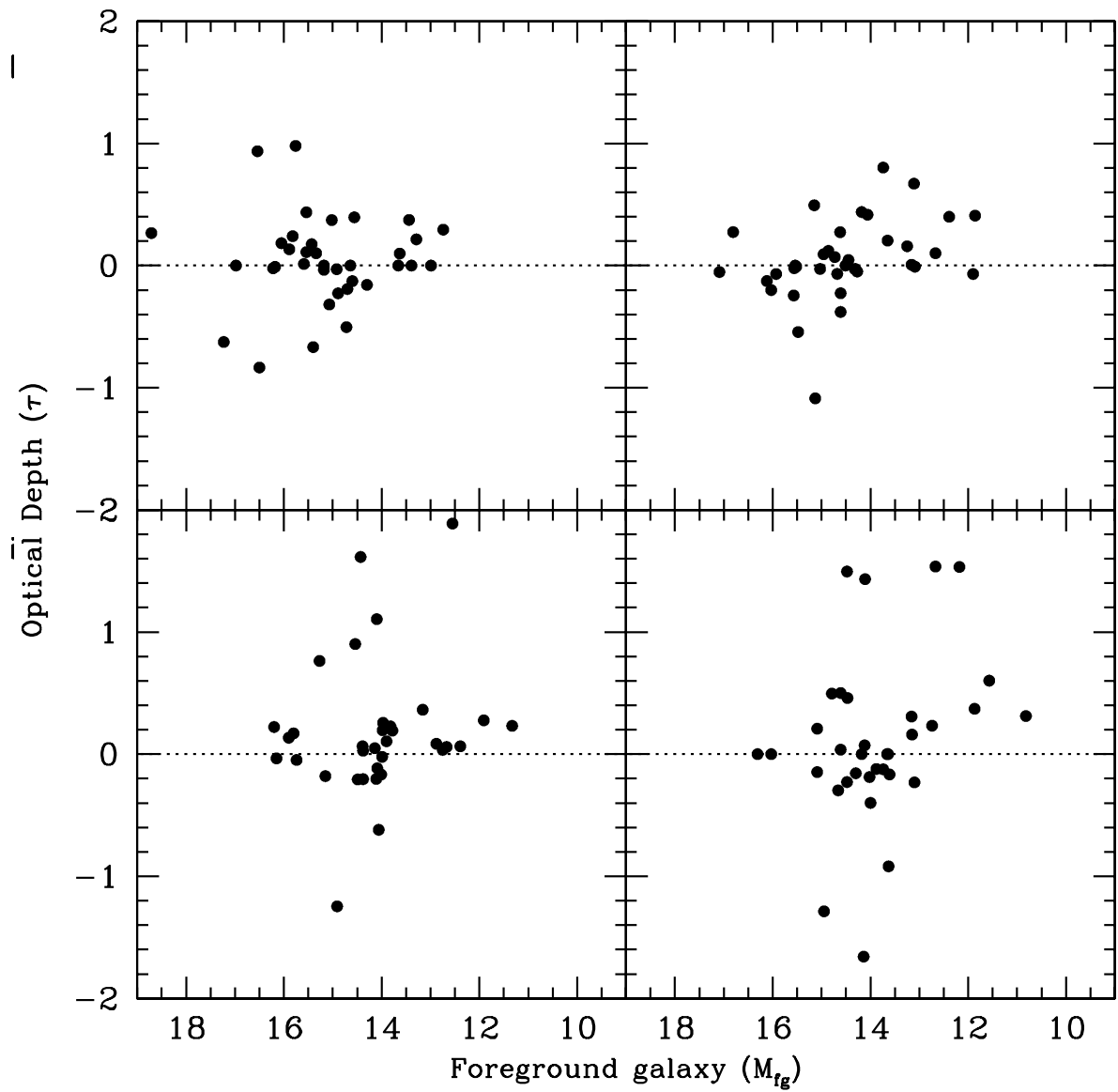}
\caption{\label{f:mt}The plot of disk optical depth and foreground galaxy magnitude --measured by the source extractor run part of the fit. There appears to be little relation between optical depth and disk brightness, probably because other affects --radial distance, arm presence and Hubble type-- dominate the optical depth value. Negative optical depth values occur when there is significant asymmetry in either or both of the members of the occulting pair.}
\end{figure}

\begin{figure}
\centering
\plotone{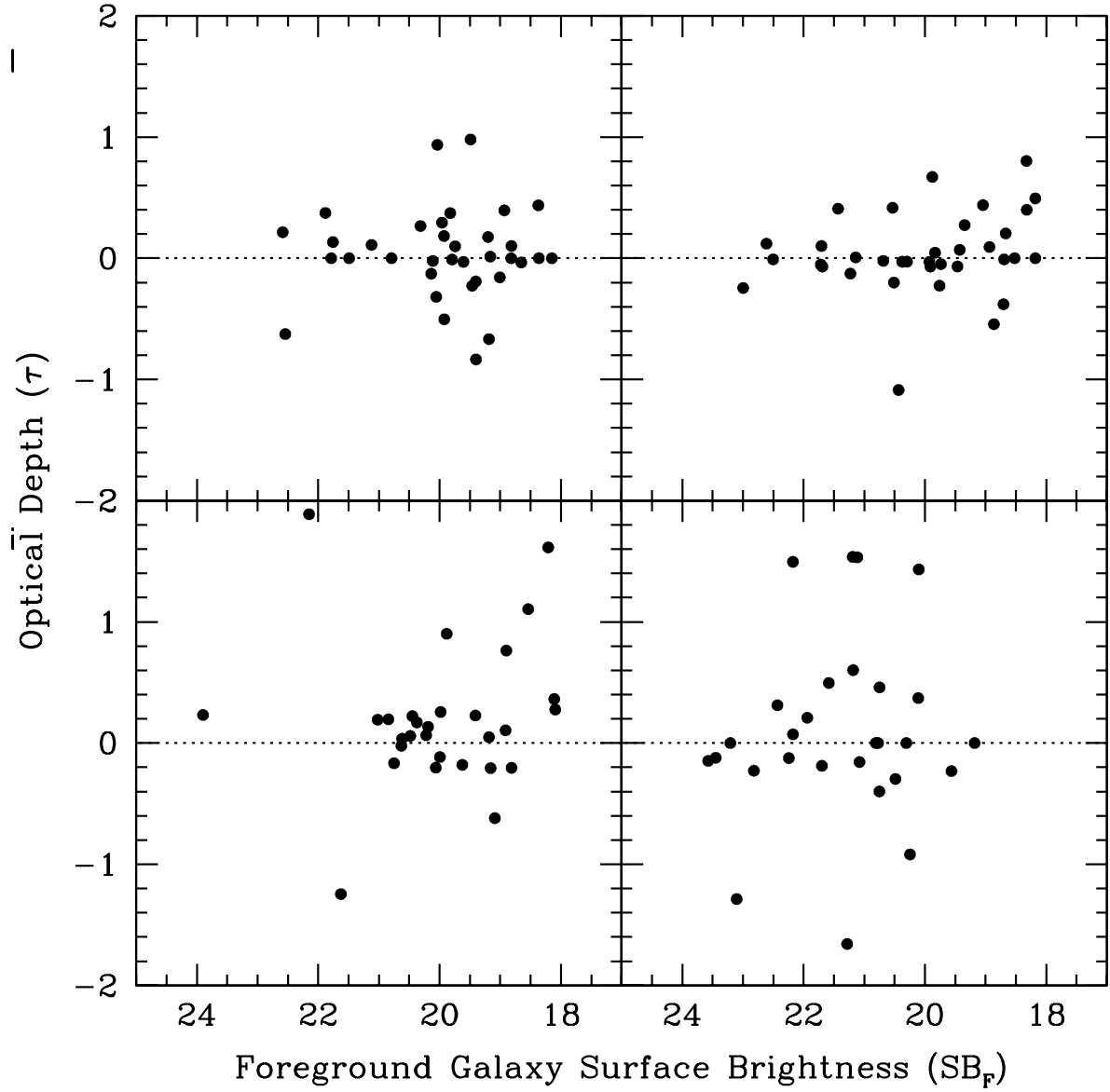}
\caption{\label{f:sbt}The plot of disk optical depth and foreground galaxy surface brightness in the aperture. No relation is evident.  }
\end{figure}

\cite{Tully98} and \cite{Masters03} note a relation between disk opacity and overall luminosity. 
A similar relation for sections of the disks between surface brightness and opacity in spiral arms is 
noted in \cite{Holwerda05b, Holwerda05d}. Figure \ref{f:mt} shows the relations between magnitude 
of the foreground galaxy and disk opacity for all the SDSS filters. No relation can be seen. Figure 
\ref{f:sbt} is the relation between optical depth and surface brightness in the visual aperture. In nearby 
galaxies there is appears to be a relation between the two \citep{Holwerda05d}, but there is little 
evidence that brighter disks are also more opaque here.

\subsection{\label{ss:R}Extinction law}

With several independent disk opacity measurements in four different filters, one could construct a 
reddening law for disks at higher redshift. However, the optical depths in the visually identified 
aperture agree with each other in all four filters (See Figure \ref{f:ra}). As noted by \cite{kw01a}, 
disk opacity is effectively gray unless sampled over disk sections smaller than 100 pc. The 
explanation is that the ISM is very clumpy, resulting in a gray extinction law when averaged 
over larger sections. 

\subsection{\label{ss:stack}Stacked images}

In order to improve the S/N of our measurement and given the expected lack of an extinction law, 
we have stacked various combinations of the SDSS filter images and fit them as a single SDSS 
filter field. In Table \ref{t:stack} and Figure \ref{f:stack}, the results are summarized. The noise 
does appear somewhat reduced in the radial profiles using r+i stacked fields (Figure \ref{f:stack}). 
Stacking more fields  together or a different combination did not result in any improvement of S/N 
(Table \ref{t:stack}).

We note that the optical depth measurements in our sample are at similar radii as those by \cite{kw00a,kw00b} 
(see Figure \ref{f:stack}). As in Figures \ref{f:ra} and \ref{f:ra:opt}, we fit an exponential profile to the 
points (parameters in Table \ref{t:ri_exp}), both to the \cite{kw00b} points and the r+i opacity points 
at different redshift. The higher redshift profile fits are extremely flat, with high scale-lengths.
If we compare the higher redshift profiles to the nearby points, they resemble the 
arm profile more than the disk. However, much of the profile is determined by opacity measurements 
nearer to the center of the spiral galaxy. 

We do not have meaningful extinction measurements close to the center of the foreground spiral 
as there is not enough flux from the background galaxy. In a deeper image, there would be significant 
flux from the background elliptical nearer to the center of the spiral, enough to make a significant 
extinction measure there. The pairs did get selected however because both fell within the 3" of the 
SDSS spectroscopic aperture. Together with the resolution to distinguish between arm and disk, 
this is another argument for follow-up with HST or possibly a large ground-based telescope.

\section{\label{s:disc}Discussion and conclusions}

The occulting galaxy technique is a well established one to measure disk extinction in the local 
universe. In combination with the wealth of data in the SDSS, we can construct a recent history 
of average disk extinction since z $\sim$ 0.3. However, due to the spatial resolution, this disk 
opacity history is for the mix or arm and disk regions for all late-type disks. The redder optical 
bands have proven themselves to be optimal for opacity measurements in these pairs due to 
the brightness of the  background early type and less asymmetry in the foreground spiral. 

\begin{figure}
\centering
\plotone{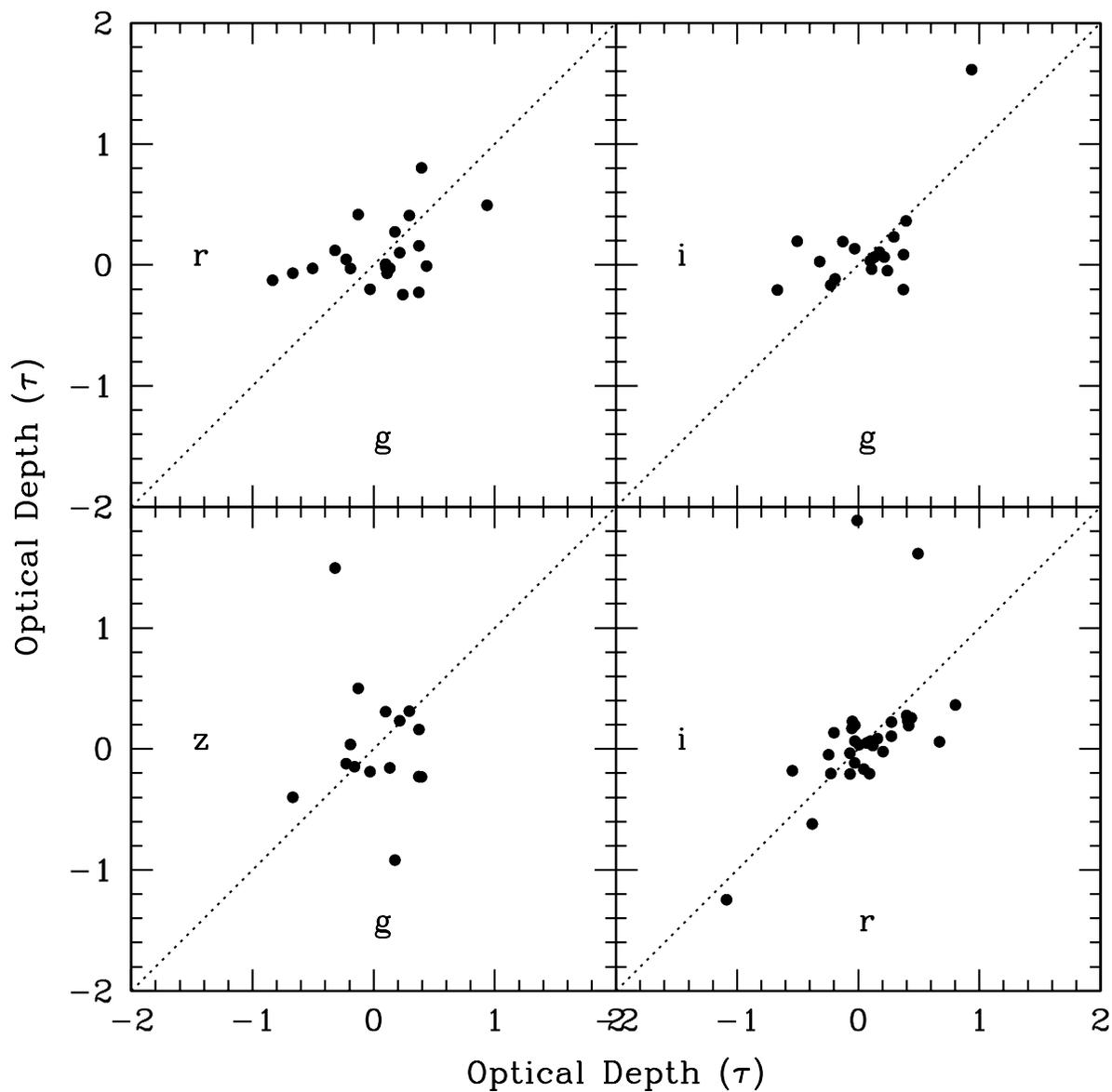}
\caption{\label{f:tt}The relation between opacity measurements in the visual aperture for four different SDSS filters ({\it g', r', i', z'}). Only the non-zero opacities are shown. The values agree for the most part for the visual aperture as a Galactic extinction law is expected to be resolved only when the spatial scale is less than 100 pc.}
\end{figure}
\begin{figure}
\centering
\plotone{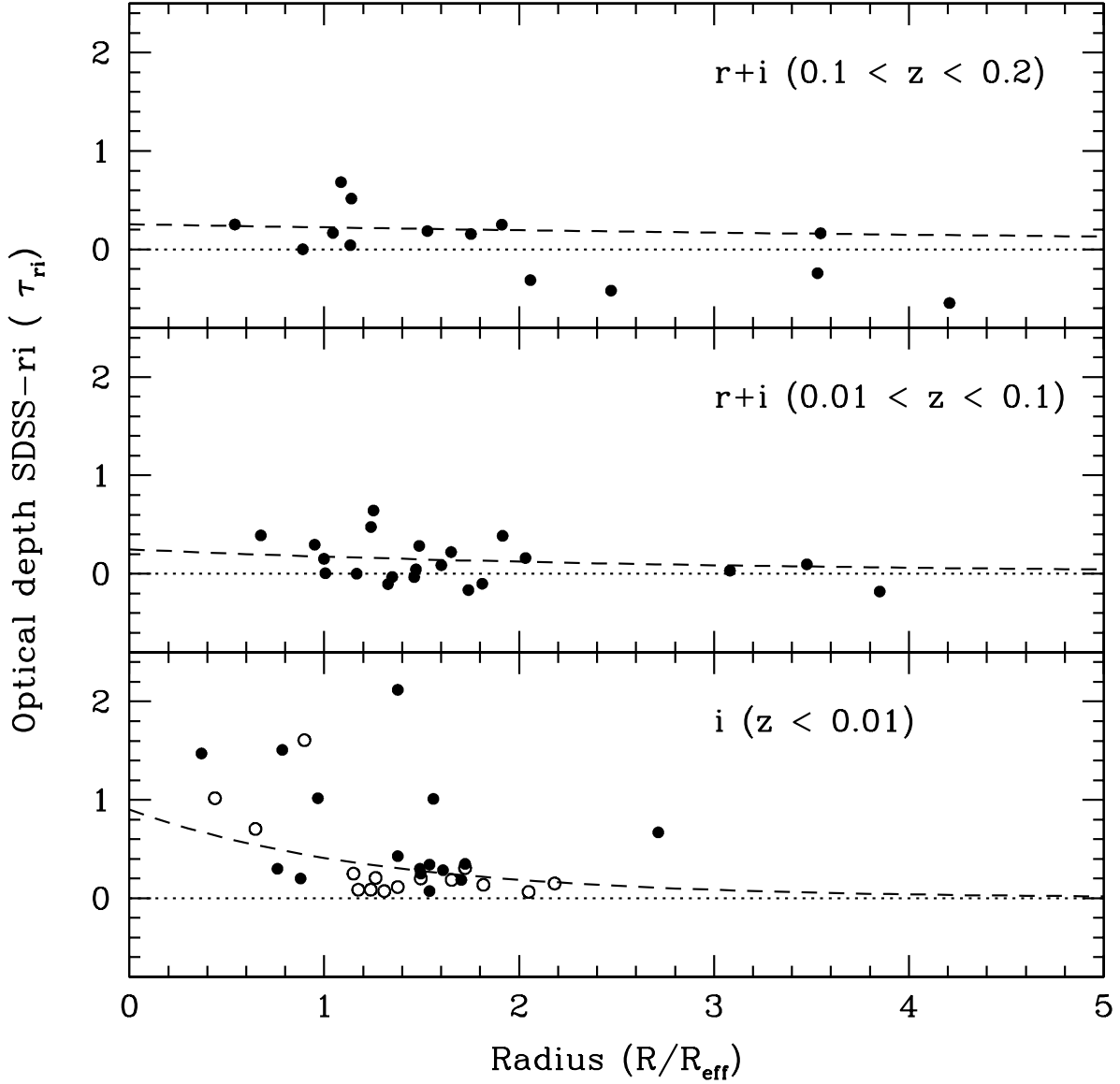}
\caption{\label{f:stack} The radial profile from Figure 14 in \cite{kw00b} (lower panel) and the optical depths profile from the r+i stacked image. The $R_{25}$ from \cite{kw00b} was converted to effective radius with $R_{25} = 1.43 \times R_{eff}$. Open circles are for disk regions, filled for spiral arm in the case of the \cite{kw00b} values. Exponential disk fits to the points are the dashed lines.}
\end{figure}

From our analysis of \sample occulting pairs in the SDSS DR4 we can conclude the following:

\begin{itemize}
\item[1.] Selection of occulting galaxy pairs from a uniform spectroscopic sample like the SDSS 
has a very high rate of success resulting in a substantial sample (\S \ref{s:sample}).
\item[2.] Fits to occulting pairs can be automated yet the optimal placement of an aperture to measure 
the disk opacity can not (\S \ref{s:method}, Figure \ref{f:ra} and \ref{f:ra:opt}).
\item[2.] The i' and r' filters in SDSS are optimal for measuring disk opacity in more distant galaxy pairs 
(Figure \ref{f:ra}).
\item[3.] The radial plots are similar to those found by \cite{kw00a} for a mix of arm and disk values (Figure \ref{f:stack}).
\item[4.] There is not enough redshift range in the current sample with successful fits to distinguish an 
evolutionary trend in disk opacity (Figure \ref{f:stack}).
\item[5.] There is no relation between overall disk luminosity and local disk opacity and only a weak 
relation between local surface brightness and disk opacity, probably because arms cannot be resolved 
at these distances by SDSS (Figures \ref{f:mt} and \ref{f:sbt}).
\item[6.] Optical depth values in different filters agree with scatter due to the different asymmetry in galaxies 
in different filters, consistent with earlier findings (Figure \ref{f:tt}).
\item[7.] Stacking SDSS images does result in some S/N improvements. In order to probe the inner parts of 
the spirals, deeper exposures are needed (Figure \ref{f:stack}).
\item[8.] The exponential profile of disk opacity for the higher redshift pairs is flatter than that of the nearby ones 
from \cite{kw00b}. More points at lower galactic radius are needed to confirm (Figure \ref{f:stack}).
\end{itemize}

\section{\label{s:fut}Future Work}

These SDSS pairs constitute the ideal sample for study of spiral disks opacity in the local Universe (z$<$0.4). 
For these and more distant pairs, the photometric stability and high resolution of HST will be needed to 1) 
resolve Hubble sub-type, 2) distinguish between arm and disk sections and allow for optical depth values 
closer to the center of the spiral. The SDSS measurements can however serve as a reference for higher-redshift 
pairs images with HST for which Hubble type and spiral arm identification also remain impossible.

\acknowledgments

The authors would like to thank Roelof de Jong, Ron Allen, and Harry Ferguson for useful discussions.  
We thank Lorraine and Steve Mencinsky for assistance in examining each candidate pair for suitable geometry.
Funding for the Sloan Digital Sky Survey (SDSS) has been provided by the Alfred P. Sloan Foundation, the 
Participating Institutions, the National Aeronautics and Space Administration, the National Science Foundation, 
the U.S. Department of Energy, the Japanese Monbukagakusho, and the Max Planck Society. The SDSS Web 
site is http://www.sdss.org/.

The SDSS is managed by the Astrophysical Research Consortium (ARC) for the Participating Institutions. 
The Participating Institutions are The University of Chicago, Fermilab, the Institute for Advanced Study, the 
Japan Participation Group, The Johns Hopkins University, Los Alamos National Laboratory, the 
Max-Planck-Institute for Astronomy (MPIA), the Max-Planck-Institute for Astrophysics (MPA), New Mexico 
State University, University of Pittsburgh, Princeton University, the United States Naval Observatory, and 
the University of Washington. W. C. Keel is grateful for support from a Dean's Leadership Board 
Faculty Fellowship.


\begin{figure}
\centering
\plotone{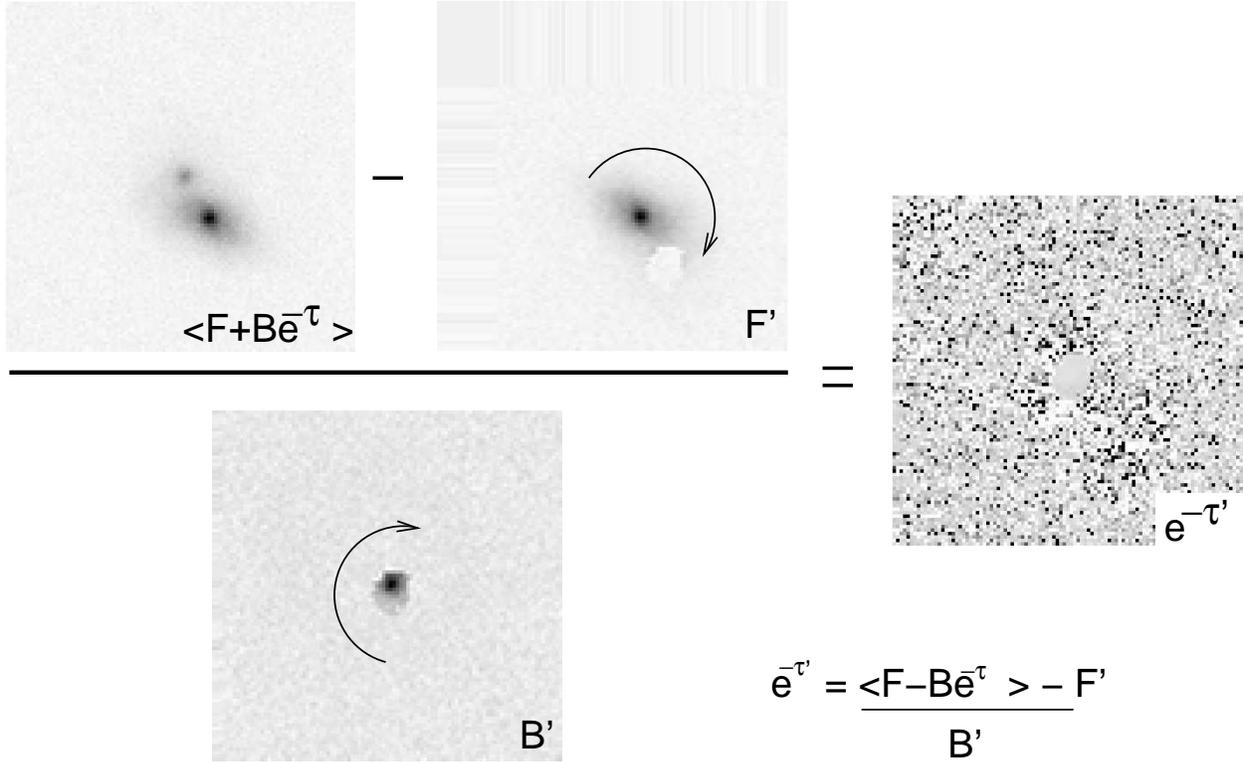}
\caption{\label{f:fit} A schematic of the automated occulting galaxy method with the pair 1006-52708-624. The extinction map is constructed when the fitparameters (central positions and rotation angles) of both pair members have been determined. The foreground galaxy is rotated and subtracted with the resulting image divided by the flipped background image. The resulting image is the opacity map. In case the area of the background image is perfectly unity, both galaxies are perfectly symmetric and there is no extinction. }
\end{figure}

\begin{figure}
\centering
\plotone{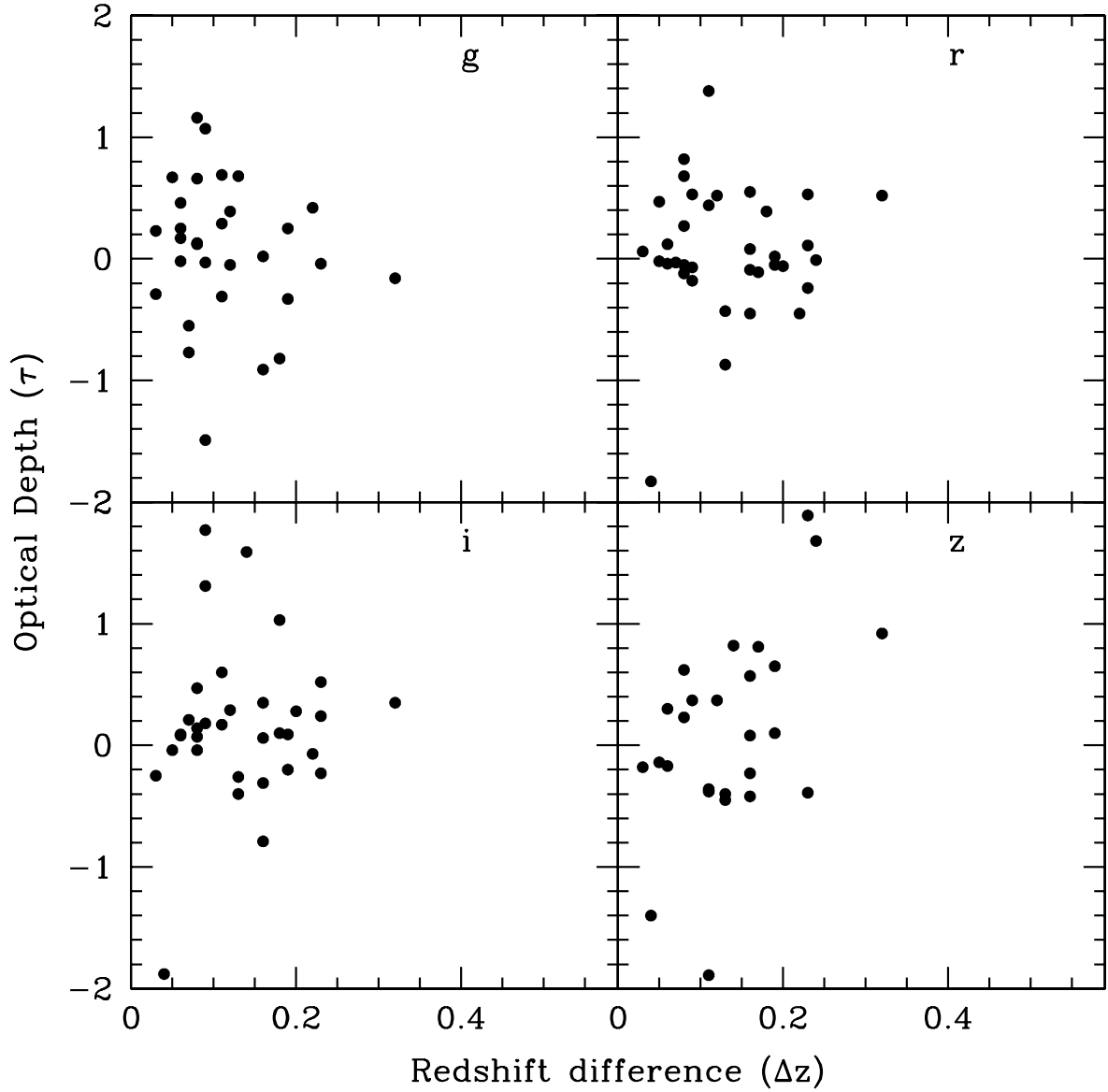}
\caption{\label{f:zzt}The plots of the redshift distance between foreground and background galaxy and the derived optical depth in the visual aperture in the SDSS for which meaningful opacities can be obtained. There appears to be no systematic relation between the distance between the pair members and the optical depth inferred.}
\end{figure}

\begin{deluxetable}{l l l l l l l l l l l l}
\tabletypesize{\footnotesize}
\tablecaption{ \label{t:sample} The SDSS basic data of the occulting galaxy pairs.}
\tablehead{ \colhead{plate}  &\colhead{mjd} & \colhead{fiber} & \colhead{ra} & \colhead{dec} & \colhead{zfg} & \colhead{zbg}  &\colhead{u} & \colhead{g} & \colhead{r} & \colhead{i} & \colhead{z}}
\startdata
       267 &      51608 &        499 & 148.311890 &   0.649980 &   0.0355 &    0.0941 &       22.79 &      20.11 &      18.79 &      18.20 &      17.66 \\ 
       273 &      51957 &        323 & 156.808510 &   0.680520 &   0.1071 &    0.1781 &       21.23 &      19.62 &      18.54 &      18.01 &      17.63 \\ 
       275 &      51910 &        233 & 161.287840 &   0.075950 &   0.0260 &    0.0944 &       21.22 &      19.10 &      18.07 &      17.54 &      17.24 \\ 
       281 &      51614 &        624 & 173.070050 &   0.608420 &   0.0468 &    0.1988 &       22.20 &      20.20 &      19.01 &      18.51 &      18.04 \\ 
       282 &      51658 &        120 & 174.666790 &  -0.146470 &   0.0722 &    0.2600 &       23.92 &      21.60 &      19.94 &      19.15 &      18.53 \\ 
       285 &      51930 &         15 & 180.269580 &  -0.787540 &   0.0213 &    0.1408 &       21.55 &      20.38 &      19.70 &      20.27 &      19.81 \\ 
       313 &      51673 &        546 & 231.835980 &   0.477040 &   0.0413 &    0.3033 &       23.41 &      21.07 &      19.43 &      18.85 &      18.42 \\ 
       367 &      51997 &        222 & 259.541780 &  55.201470 &   0.0836 &    0.1394 &       21.30 &      19.50 &      18.44 &      17.94 &      17.54 \\ 
       403 &      51871 &        586 &  30.145580 &   0.560980 &   0.1378 &    0.1658 &       21.70 &      19.71 &      18.58 &      18.04 &      17.66 \\ 
       408 &      51821 &        336 &  38.992950 &   1.257390 &   0.0682 &    0.1243 &       23.31 &      20.47 &      19.34 &      18.65 &      18.23 \\ 
       435 &      51882 &        118 & 118.882910 &  40.503790 &   0.2313 &    0.2620 &       23.54 &      21.07 &      19.45 &      18.82 &      18.32 \\ 
       436 &      51883 &        638 & 120.747080 &  45.492670 &   0.0389 &    0.3643 &       23.75 &      22.78 &      20.48 &      19.75 &      19.28 \\ 
       440 &      51885 &        370 & 123.106550 &  50.404870 &   0.1756 &    0.2453 &       23.91 &      21.72 &      20.09 &      19.24 &      18.93 \\ 
       464 &      51908 &        211 &  58.942650 &  -5.948270 &   0.1230 &    0.2011 &       24.24 &      20.67 &      19.37 &      18.68 &      18.22 \\ 
       474 &      52000 &        548 & 142.017490 &   0.242640 &   0.2061 &    0.2970 &       22.94 &      20.72 &      19.15 &      18.58 &      18.21 \\ 
       484 &      51907 &        616 & 139.814360 &  59.294570 &   0.1582 &    0.2453 &       25.28 &      21.65 &      19.81 &      19.15 &      18.83 \\ 
       487 &      51943 &        305 & 147.228910 &  62.483730 &   0.1245 &    0.2479 &       24.03 &      21.38 &      19.78 &      19.09 &      18.70 \\ 
       497 &      51989 &         73 & 208.034660 &  65.742370 &   0.1953 &    0.3958 &       23.67 &      21.82 &      20.06 &      19.23 &      18.63 \\ 
       499 &      51988 &        152 & 217.024030 &  63.613230 &   0.1145 &    0.2863 &       25.86 &      22.30 &      20.05 &      19.48 &      19.10 \\ 
       511 &      52636 &        313 & 169.186260 &   2.065220 &   0.1292 &    0.2673 &       22.10 &      20.43 &      19.16 &      18.52 &      18.12 \\ 
       512 &      51992 &        302 & 171.158050 &   1.820180 &   0.0492 &    0.2321 &       22.85 &      20.44 &      18.90 &      18.32 &      17.94 \\ 
       529 &      52025 &        408 & 205.440810 &   3.431220 &   0.0232 &    0.1681 &       21.74 &      19.97 &      18.92 &      18.45 &      18.14 \\ 
       545 &      52202 &        504 & 121.639780 &  40.920020 &   0.0764 &    0.3170 &       24.52 &      22.43 &      20.28 &      19.58 &      19.19 \\ 
       547 &      52207 &        167 & 124.385840 &  43.503210 &   0.1415 &    0.3761 &       24.73 &      22.51 &      20.37 &      19.56 &      19.05 \\ 
       550 &      51959 &        624 & 131.782940 &  48.626320 &   0.1743 &    0.1975 &       24.74 &      21.65 &      19.57 &      18.89 &      18.57 \\ 
       628 &      52083 &        262 & 249.181810 &  41.790430 &   0.0284 &    0.1203 &       21.19 &      19.49 &      18.55 &      18.18 &      17.75 \\ 
       629 &      52051 &         26 & 252.894690 &  43.849070 &   0.1152 &    0.3502 &       23.01 &      22.07 &      20.40 &      19.41 &      19.08 \\ 
       637 &      52174 &        334 & 314.058170 &  -6.743460 &   0.0878 &    0.1779 &       23.42 &      20.95 &      19.29 &      18.65 &      18.31 \\ 
       671 &      52206 &        347 &  34.784680 &   0.615480 &   0.0408 &    0.2803 &       22.02 &      21.57 &      20.60 &      20.28 &      19.76 \\ 
       765 &      52254 &        179 & 137.082920 &  49.082730 &   0.0355 &    0.2128 &       21.32 &      19.83 &      18.68 &      18.18 &      17.80 \\ 
       782 &      52320 &        306 & 190.878970 &  62.349390 &   0.1452 &    0.2056 &       22.21 &      20.15 &      18.82 &      18.25 &      17.87 \\ 
       790 &      52441 &        527 & 219.458290 &  58.911160 &   0.0317 &    0.1381 &       22.33 &      19.79 &      18.59 &      18.08 &      17.81 \\ 
       792 &      52353 &        432 & 224.462030 &  56.091590 &   0.1423 &    0.2012 &       22.25 &      20.27 &      18.77 &      18.19 &      17.85 \\ 
       808 &      52556 &         39 &  46.965700 &  -0.311940 &   0.2139 &    0.3731 &       23.21 &      21.72 &      20.34 &      19.66 &      19.27 \\ 
       845 &      52381 &        270 & 187.675090 &   4.707290 &   0.0700 &    0.3164 &       24.73 &      21.85 &      19.52 &      18.88 &      18.60 \\ 
       848 &      52669 &        602 & 195.053830 &   5.739970 &   0.0490 &    0.2219 &       22.22 &      20.39 &      19.20 &      18.80 &      18.39 \\ 
       860 &      52319 &        451 & 121.339030 &  30.762010 &   0.0370 &    0.1607 &       21.94 &      21.15 &      18.59 &      17.94 &      17.59 \\ 
       864 &      52320 &         91 & 129.972990 &  35.742270 &   0.0439 &    0.2576 &       22.17 &      20.78 &      19.72 &      19.18 &      18.83 \\ 
       877 &      52353 &        458 & 166.000290 &  51.786350 &   0.0370 &    0.0616 &       21.28 &      20.11 &      19.54 &      19.19 &      19.01 \\ 
       882 &      52370 &        122 & 180.578770 &  51.309050 &   0.0599 &    0.1828 &       21.51 &      19.41 &      18.26 &      17.76 &      17.45 \\ 
       883 &      52430 &        366 & 182.350150 &  53.640160 &   0.0492 &    0.1319 &       21.68 &      19.73 &      18.65 &      18.12 &      17.73 \\ 
       884 &      52374 &        242 & 186.059060 &  51.293770 &   0.0413 &    0.1745 &       21.35 &      19.56 &      18.42 &      17.94 &      17.62 \\ 
       889 &      52663 &        408 & 116.692660 &  30.657980 &   0.0558 &    0.1612 &       21.44 &      19.77 &      18.57 &      18.02 &      17.63 \\ 
       896 &      52592 &        463 & 131.060470 &  43.308520 &   0.0275 &    0.1059 &       21.57 &      19.52 &      18.41 &      17.96 &      17.56 \\ 
       913 &      52433 &        151 & 206.813230 &  -2.646100 &   0.0843 &    0.1583 &       21.77 &      19.58 &      18.32 &      17.79 &      17.38 \\ 
       931 &      52619 &         84 & 124.763000 &  29.720850 &   0.1835 &    0.2781 &       22.60 &      20.79 &      19.32 &      18.68 &      18.28 \\ 
       958 &      52410 &        194 & 197.246590 &  59.181390 &   0.1523 &    0.3145 &       25.22 &      21.71 &      20.00 &      19.28 &      19.05 \\ 
       970 &      52413 &        408 & 183.318710 &  51.174320 &   0.3046 &    0.3819 &       22.92 &      21.27 &      19.78 &      19.08 &      18.60 \\ 
       973 &      52426 &        519 & 255.808640 &  33.511540 &   0.0631 &    0.0914 &       21.88 &      19.07 &      18.00 &      17.47 &      17.15 \\ 
       977 &      52410 &        539 & 258.531050 &  29.864680 &   0.0824 &    0.1954 &       22.13 &      19.91 &      18.67 &      18.11 &      17.80 \\ 
       980 &      52431 &        300 & 258.937370 &  27.421190 &   0.1253 &    0.2907 &       24.22 &      21.76 &      19.48 &      18.78 &      18.44 \\ 
      1000 &      52643 &        337 & 159.372270 &   7.579740 &   0.0851 &    0.3245 &       24.23 &      21.53 &      19.44 &      18.83 &      18.51 \\ 
      1006 &      52708 &        624 & 152.436350 &  49.838300 &   0.0524 &    0.1323 &       21.96 &      19.99 &      18.47 &      18.02 &      17.50 \\ 
      1007 &      52706 &        519 & 153.131040 &  50.738660 &   0.0463 &    0.1952 &       21.60 &      19.74 &      18.47 &      17.93 &      17.63 \\ 
      1160 &      52674 &        468 & 214.447800 &  56.804440 &   0.2381 &    0.2914 &       22.28 &      20.80 &      19.58 &      19.12 &      18.81 \\ 
      1215 &      52725 &        288 & 143.281310 &  39.220680 &   0.2708 &    0.3340 &       22.71 &      21.18 &      19.56 &      18.95 &      18.60 \\ 
      1230 &      52672 &        639 & 185.753290 &  10.099700 &   0.0536 &    0.1608 &       21.67 &      20.04 &      19.04 &      18.57 &      18.17 \\ 
      1235 &      52734 &        164 & 148.326780 &   7.600730 &   0.0956 &    0.3255 &       22.21 &      20.84 &      19.34 &      18.73 &      18.33 \\ 
      1269 &      52937 &        485 & 130.654940 &  30.422070 &   0.0272 &    0.1878 &       21.30 &      19.90 &      18.91 &      18.48 &      18.23 \\ 
      1280 &      52738 &        125 & 195.231990 &  48.583740 &   0.0799 &    0.2089 &       21.89 &      20.13 &      18.95 &      18.45 &      18.14 \\ 
      1282 &      52759 &        630 & 202.872560 &  49.243360 &   0.1245 &    0.2786 &       21.92 &      20.42 &      19.06 &      18.51 &      18.14 \\ 
      1310 &      53033 &        202 & 173.008190 &  56.231840 &   0.1140 &    0.1840 &       21.76 &      19.95 &      18.75 &      18.26 &      17.89 \\ 
      1317 &      52765 &          9 & 192.781290 &  56.429170 &   0.0974 &    0.2008 &       21.59 &      19.86 &      18.60 &      18.06 &      17.64 \\ 
      1321 &      52764 &        461 & 203.989570 &  56.632780 &   0.1528 &    0.2173 &       22.30 &      20.49 &      19.11 &      18.56 &      18.24 \\ 
      1324 &      53088 &         50 & 212.489360 &  53.497750 &   0.0422 &    0.2251 &       22.31 &      20.52 &      19.31 &      18.79 &      18.46 \\ 
      1325 &      52762 &        313 & 211.388780 &  54.126820 &   0.1168 &    0.2119 &       22.14 &      20.29 &      18.96 &      18.48 &      18.10 \\ 
      1327 &      52781 &        252 & 218.738380 &  51.467310 &   0.0799 &    0.1408 &       21.41 &      19.79 &      18.77 &      18.31 &      17.97 \\ 
      1332 &      52781 &        469 & 233.667240 &  46.404020 &   0.2849 &    0.3199 &       22.41 &      20.89 &      19.44 &      18.89 &      18.57 \\ 
      1363 &      53053 &        104 & 166.400210 &  42.239340 &   0.1759 &    0.2673 &       22.44 &      20.71 &      19.28 &      18.73 &      18.41 \\ 
      1373 &      53063 &        255 & 191.929650 &  43.917920 &   0.0294 &    0.1317 &       21.48 &      19.63 &      18.61 &      18.20 &      17.87 \\ 
      1388 &      53119 &        618 & 233.838020 &  32.213560 &   0.0370 &    0.2180 &       21.91 &      20.30 &      19.18 &      18.69 &      18.36 \\ 
      1390 &      53142 &        385 & 236.455400 &  29.691260 &   0.1313 &    0.2863 &       22.17 &      20.44 &      19.02 &      18.42 &      18.02 \\ 
      1402 &      52872 &        554 & 236.160490 &  35.176090 &   0.0550 &    0.0790 &       21.02 &      19.26 &      18.35 &      17.90 &      17.60 \\ 
      1416 &      52875 &        490 & 236.353870 &  36.378000 &   0.0653 &    0.1909 &       21.48 &      20.05 &      19.03 &      18.56 &      18.17 \\ 
      1423 &      53167 &        528 & 250.832630 &  26.703910 &   0.1975 &    0.3322 &       22.71 &      21.35 &      19.89 &      19.30 &      18.92 \\ 
      1426 &      52993 &        385 & 151.191870 &  37.772590 &   0.0232 &    0.0519 &       20.39 &      18.62 &      17.78 &      17.40 &      17.09 \\ 
      1429 &      52990 &        336 & 156.275830 &  45.059290 &   0.0742 &    0.1461 &       21.44 &      19.68 &      18.50 &      17.97 &      17.62 \\ 
      1604 &      53078 &        459 & 168.059410 &  12.811400 &   0.0771 &    0.1681 &       21.30 &      19.22 &      18.03 &      17.54 &      17.15 \\ 
      1607 &      53083 &         48 & 174.721650 &  11.288210 &   0.0814 &    0.1513 &       21.59 &      19.93 &      18.80 &      18.42 &      18.00 \\ 
      1618 &      53116 &        285 & 171.000090 &   6.105510 &   0.0370 &    0.1741 &       21.50 &      19.68 &      18.56 &      18.11 &      17.74 \\ 
      1620 &      53137 &        175 & 175.693770 &   7.235100 &   0.0653 &    0.1011 &       21.21 &      19.36 &      18.38 &      17.89 &      17.50 \\ 
      1677 &      53148 &        262 & 226.092130 &  43.630090 &   0.1931 &    0.3080 &       22.36 &      20.87 &      19.45 &      18.88 &      18.46 \\ 
      1746 &      53062 &        533 & 155.484860 &  14.117880 &   0.1402 &    0.2469 &       22.03 &      20.37 &      19.02 &      18.46 &      18.12 \\       
\enddata
\end{deluxetable}

\begin{deluxetable}{l l l l l l l l l l l l l}
\tabletypesize{\footnotesize}
\tablewidth{0pt.}
\tablecaption{ \label{t:tau} The optical depths of the SDSS filters. These values have not been corrected for inclination.}
\tablehead{ \colhead{plate}  &\colhead{mjd} & \colhead{fiber} & \colhead{$z_{fg}$} & \colhead{$z_{bg}$} & \colhead{R} & \colhead{Incl} & \colhead{$\tau_u$} &\colhead{$\tau_g$} &\colhead{$\tau_r$} &\colhead{$\tau_i$} &\colhead{$\tau_z$} }
\startdata
   267 &  51608 &    499 &   0.04 &   0.09 &   \dots (\dots) &    0.0 &   \dots (\dots) &    0.44 (  3.65) &   -0.01 ( 26.35) &   \dots (\dots) &   \dots (  1.84) \\ 
   273 &  51957 &    323 &   0.11 &   0.18 &   \dots (\dots) &    0.0 &   \dots (\dots) &    \dots (\dots) &    \dots (\dots) &   \dots (\dots) &   \dots (\dots) \\ 
   275 &  51910 &    233 &   0.03 &   0.09 &   0.81 (  0.13) &   39.0 &  -0.39 (  2.39) &    0.21 (  3.61) &    0.10 ( 11.02) &   0.06 (  2.14) &   0.23 (  1.79) \\ 
   281 &  51614 &    624 &   0.05 &   0.20 &   \dots (\dots) &    0.0 &   \dots (\dots) &    \dots (\dots) &    \dots (\dots) &   \dots (\dots) &   \dots (\dots) \\ 
   282 &  51658 &    120 &   0.07 &   0.26 &   1.82 (  0.27) &   61.7 &   \dots (\dots) &    0.10 (  2.19) &    0.01 (  3.85) &   0.04 (  5.93) &   0.31 (  1.82) \\ 
   285 &  51930 &     15 &   0.02 &   0.14 &   \dots (\dots) &    0.0 &   \dots (\dots) &    \dots (\dots) &    \dots (\dots) &   \dots (\dots) &   0.0 ( 18.51) \\ 
   313 &  51673 &    546 &   0.04 &   0.30 &   \dots (\dots) &    0.0 &   \dots (\dots) &   -4.76 (  1.94) &    \dots (\dots) &   \dots (\dots) &   \dots (\dots) \\ 
   367 &  51997 &    222 &   0.08 &   0.14 &   \dots (\dots) &    0.0 &   \dots (\dots) &    \dots (\dots) &    \dots (\dots) &   \dots (\dots) &   0.0 (  2.63) \\ 
   403 &  51871 &    586 &   0.14 &   0.17 &   \dots (\dots) &    0.0 &   \dots (\dots) &    \dots (\dots) &    \dots (\dots) &   \dots (\dots) &   0.0 (  2.29) \\ 
   408 &  51821 &    336 &   0.07 &   0.12 &   1.11 (  0.45) &   27.7 &   \dots (\dots) &    \dots (\dots) &    0.20 (  4.11) &  -0.02 (  1.86) &  -0.12 (  2.78) \\ 
   435 &  51882 &    118 &   0.23 &   0.26 &   \dots (\dots) &    0.0 &   \dots (\dots) &    \dots (\dots) &    \dots (\dots) &   \dots (\dots) &   0.0 (  1.55) \\ 
   436 &  51883 &    638 &   0.04 &   0.36 &   1.63 (  0.27) &   57.0 &   \dots (\dots) &   -0.13 (  4.92) &    0.42 (  2.31) &   0.19 (  1.88) &   0.50 (  2.66) \\ 
   440 &  51885 &    370 &   0.18 &   0.25 &   \dots (\dots) &    0.0 &   \dots (\dots) &    \dots (\dots) &    \dots (\dots) &   \dots (\dots) &   0.0 (  1.67) \\ 
   464 &  51908 &    211 &   0.12 &   0.20 &   \dots (\dots) &    0.0 &   \dots (\dots) &    0.98 (  3.12) &    \dots (\dots) &   \dots (\dots) &   \dots (\dots) \\ 
   474 &  52000 &    548 &   0.21 &   0.30 &   \dots (\dots) &    0.0 &   \dots (\dots) &    \dots (\dots) &    \dots (\dots) &   \dots (\dots) &   \dots (\dots) \\ 
   484 &  51907 &    616 &   0.16 &   0.25 &   \dots (\dots) &    0.0 &   \dots (\dots) &   -0.83 (  6.27) &   -0.13 (  2.13) &   \dots (\dots) &   \dots (\dots) \\ 
   487 &  51943 &    305 &   0.12 &   0.25 &   3.48 (  0.39) &   59.5 &   \dots (\dots) &    0.37 (  6.54) &   -0.23 (  7.13) &  -0.20 (  5.69) &  -0.23 (\dots) \\ 
   497 &  51989 &     73 &   0.20 &   0.40 &   1.12 (  0.42) &   38.6 &   \dots (\dots) &    \dots (\dots) &   -0.05 ( 24.50) &   0.23 (  2.12) &  -1.66 (\dots) \\ 
   499 &  51988 &    152 &   0.11 &   0.29 &   1.81 (  0.64) &   49.7 &   \dots (\dots) &    \dots (\dots) &    0.0 (  2.54) &   0.76 ( 61.34) &   0.0 (  1.50) \\ 
   511 &  52636 &    313 &   0.13 &   0.27 &   0.81 (  0.37) &   52.8 &   \dots (\dots) &    \dots (\dots) &    \dots (\dots) &   0.90 (  4.77) &   0.50 (  2.29) \\ 
   512 &  51992 &    302 &   0.05 &   0.23 &   1.26 (  0.29) &   44.3 &   \dots (\dots) &   -0.32 (  1.72) &    0.12 (  2.86) &   0.03 (  3.82) &   1.50 (  5.58) \\ 
   529 &  52025 &    408 &   0.02 &   0.17 &   \dots (\dots) &    0.0 &   \dots (\dots) &    \dots (\dots) &    \dots (\dots) &   \dots (\dots) &   0.0 (  1.70) \\ 
   550 &  51959 &    624 &   0.17 &   0.20 &   \dots (\dots) &    0.0 &   \dots (\dots) &    0.18 ( 78.96) &    \dots (\dots) &   \dots (\dots) &   \dots (\dots) \\ 
   628 &  52083 &    262 &   0.03 &   0.12 &   \dots (\dots) &    0.0 &   \dots (\dots) &    \dots (\dots) &    \dots (\dots) &   \dots (\dots) &   0.0 (  2.31) \\ 
   629 &  52051 &     26 &   0.12 &   0.35 &   0.85 (  0.34) &   40.7 &   \dots (\dots) &    \dots (\dots) &    0.09 (  2.35) &  -0.20 (  2.18) &   1.43 ( 26.06) \\ 
   637 &  52174 &    334 &   0.09 &   0.18 &   1.06 (  0.46) &   48.6 &   \dots (\dots) &    0.0 (  1.67) &    0.0 (  4.10) &   1.10 (  2.86) &   0.0 (  3.29) \\ 
   671 &  52206 &    347 &   0.04 &   0.28 &   4.23 (  0.14) &   24.2 &   \dots (\dots) &    0.0 (  6.47) &   -0.01 (  2.07) &   1.89 ( 13.63) &   1.53 (\dots) \\ 
   765 &  52254 &    179 &   0.04 &   0.21 &   \dots (\dots) &    0.0 &   \dots (\dots) &    \dots (\dots) &    \dots (\dots) &   \dots (\dots) &   \dots (\dots) \\ 
   782 &  52320 &    306 &   0.15 &   0.21 &   \dots (\dots) &    0.0 &   \dots (\dots) &    0.27 (  1.36) &    \dots (\dots) &   \dots (\dots) &   \dots (\dots) \\ 
   790 &  52441 &    527 &   0.03 &   0.14 &   1.83 (  0.30) &   67.3 &   0.0 (  2.24) &   -0.16 (  4.21) &    \dots (\dots) &   \dots (\dots) &  -0.15 (  2.43) \\ 
   792 &  52353 &    432 &   0.14 &   0.20 &   1.76 (  0.41) &   23.1 &   \dots (\dots) &    0.13 (  2.70) &   -0.03 (  2.03) &   0.06 (  2.04) &  -0.16 (  1.81) \\ 
   808 &  52556 &     39 &   0.21 &   0.37 &   \dots (\dots) &    0.0 &   \dots (\dots) &    \dots (\dots) &    \dots (\dots) &   \dots (\dots) &   \dots (\dots) \\ 
   845 &  52381 &    270 &   0.07 &   0.32 &   \dots (\dots) &    0.0 &   \dots (\dots) &    \dots (\dots) &    \dots (\dots) &   \dots (\dots) &   \dots (\dots) \\ 
   848 &  52669 &    602 &   0.05 &   0.22 &   2.22 (  0.19) &   42.0 &   \dots (\dots) &    \dots (\dots) &   -0.07 (  1.91) &   \dots (\dots) &   0.60 (  2.64) \\ 
   860 &  52319 &    451 &   0.04 &   0.16 &   1.66 (  0.23) &   32.5 &   0.0 (  2.33) &    0.29 (  0.79) &    0.41 (  1.02) &   0.23 (  2.62) &   0.31 (  2.04) \\ 
   864 &  52320 &     91 &   0.04 &   0.26 &   \dots (\dots) &    0.0 &   \dots (\dots) &    0.24 (  1.81) &   -0.24 ( 19.63) &  -0.05 (  2.01) &   \dots (\dots) \\ 
   877 &  52353 &    458 &   0.04 &   0.06 &   \dots (\dots) &    0.0 &   \dots (\dots) &    \dots (\dots) &    \dots (\dots) &   \dots (\dots) &   \dots (\dots) \\ 
   882 &  52370 &    122 &   0.06 &   0.18 &   \dots (\dots) &    0.0 &   \dots (\dots) &    \dots (\dots) &    \dots (\dots) &   \dots (\dots) &   0.0 ( 68.77) \\ 
   883 &  52430 &    366 &   0.05 &   0.13 &   \dots (\dots) &    0.0 &   \dots (\dots) &    0.11 (  4.58) &   -0.07 (  2.56) &  -0.03 (  6.10) &   \dots (\dots) \\ 
   884 &  52374 &    242 &   0.04 &   0.17 &   \dots (\dots) &    0.0 &   \dots (\dots) &    \dots (\dots) &    \dots (\dots) &   \dots (\dots) &   \dots (  3.84) \\ 
   889 &  52663 &    408 &   0.06 &   0.16 &   \dots (\dots) &    0.0 &   \dots (\dots) &    \dots (\dots) &    \dots (\dots) &   \dots (\dots) &   \dots (\dots) \\ 
   896 &  52592 &    463 &   0.03 &   0.11 &   2.49 (  0.19) &   45.8 &   0.0 (123.79) &    0.37 ( 10.70) &    0.16 (  5.94) &   0.09 ( 47.10) &   0.16 (\dots) \\ 
   913 &  52433 &    151 &   0.08 &   0.16 &   \dots (\dots) &    0.0 &   \dots (\dots) &    0.10 (  1.79) &   -0.02 (  2.56) &   \dots (\dots) &   \dots (  2.36) \\ 
   931 &  52619 &     84 &   0.18 &   0.28 &   \dots (\dots) &    0.0 &   \dots (\dots) &    \dots (\dots) &    \dots (\dots) &   \dots (\dots) &   \dots (\dots) \\ 
   958 &  52410 &    194 &   0.15 &   0.31 &   4.49 (  0.81) &   36.3 &   \dots (\dots) &    \dots (\dots) &   -0.38 (  2.72) &  -0.62 (  1.29) &   0.46 (\dots) \\ 
   970 &  52413 &    408 &   0.30 &   0.38 &   \dots (\dots) &    0.0 &   \dots (\dots) &    \dots (\dots) &    \dots (\dots) &   \dots (\dots) &   \dots (  3.52) \\ 
   980 &  52431 &    300 &   0.13 &   0.29 &   1.49 (  0.20) &   18.1 &   \dots (\dots) &   -0.67 (  2.25) &   -0.07 (  1.48) &  -0.21 (  1.76) &  -0.40 (\dots) \\ 
  1000 &  52643 &    337 &   0.09 &   0.32 &   \dots (\dots) &    0.0 &   \dots (\dots) &    \dots (\dots) &    0.27 (  2.88) &   0.22 (1860.97) &   \dots (\dots) \\ 
  1006 &  52708 &    624 &   0.05 &   0.13 &   1.86 (  0.24) &   53.2 &   \dots (\dots) &    0.0 (  1.54) &    0.40 (  1.60) &   0.28 (  2.76) &   0.37 (  1.88) \\ 
  1007 &  52706 &    519 &   0.05 &   0.20 &   \dots (\dots) &    0.0 &   \dots (\dots) &    \dots (\dots) &    \dots (\dots) &   \dots (\dots) &   \dots (\dots) \\ 
  1160 &  52674 &    468 &   0.24 &   0.29 &   \dots (\dots) &    0.0 &   \dots (\dots) &    \dots (\dots) &    \dots (\dots) &   \dots (\dots) &   0.0 (  4.23) \\ 
  1215 &  52725 &    288 &   0.27 &   0.33 &   \dots (\dots) &    0.0 &   \dots (\dots) &   -0.01 (  1.73) &    \dots (\dots) &   \dots (\dots) &   0.0 (  1.96) \\ 
  1230 &  52672 &    639 &   0.05 &   0.16 &   1.19 (  0.37) &   60.9 &   0.0 (  2.30) &    0.18 (  2.21) &    0.27 (  2.85) &   0.11 (125.74) &  -0.92 (\dots) \\ 
  1235 &  52734 &    164 &   0.10 &   0.33 &   0.88 (  0.24) &   61.2 &   \dots (\dots) &   -0.03 (  1.94) &   -0.20 (  3.08) &   0.13 (  9.64) &  -0.19 (\dots) \\ 
  1269 &  52937 &    485 &   0.03 &   0.19 &   \dots (\dots) &    0.0 &   \dots (\dots) &    0.01 (  5.78) &    \dots (\dots) &   \dots (\dots) &   \dots (\dots) \\ 
  1280 &  52738 &    125 &   0.08 &   0.21 &   \dots (\dots) &    0.0 &   \dots (\dots) &    \dots (\dots) &    \dots (\dots) &   \dots (\dots) &   \dots (\dots) \\ 
  1282 &  52759 &    630 &   0.12 &   0.28 &   0.57 (  0.26) &   43.7 &   \dots (\dots) &    0.0 (  1.93) &    0.44 (207.10) &   0.26 ( 19.18) &  -0.17 (  1.72) \\ 
  1310 &  53033 &    202 &   0.11 &   0.18 &   \dots (\dots) &    0.0 &   \dots (\dots) &    \dots (\dots) &    \dots (\dots) &   \dots (\dots) &   \dots (\dots) \\ 
  1317 &  52765 &      9 &   0.10 &   0.20 &   \dots (\dots) &    0.0 &   \dots (\dots)   &    \dots (\dots) &    \dots (\dots) &   \dots (\dots) &   \dots (\dots) \\ 
  1321 &  52764 &    461 &   0.15 &   0.22 &   2.14 (  0.41) &   35.4 &   \dots (\dots) &   -0.50 ( 32.41) &   -0.03 (  3.15) &   0.20 (  1.73) &   \dots (\dots) \\ 
  1324 &  53088 &     50 &   0.04 &   0.23 &   1.97 (  0.32) &   68.4 &   \dots (\dots)  &   -0.19 (  2.20) &   -0.03 (  2.16) &  -0.12 (  2.98) &   0.04 (\dots) \\ 
  1325 &  52762 &    313 &   0.12 &   0.21 &   \dots (\dots) &    0.0 &   \dots (\dots) &   -0.02 (  2.39) &    \dots (\dots) &   \dots (\dots) &   \dots (\dots) \\ 
  1327 &  52781 &    252 &   0.08 &   0.14 &   \dots (\dots) &    0.0 &   \dots (\dots) &    \dots (\dots) &    \dots (\dots) &   \dots (\dots) &   \dots (\dots) \\ 
  1332 &  52781 &    469 &   0.28 &   0.32 &   5.83 (  0.64) &   23.2 &   \dots (\dots) &    \dots (\dots) &   -1.09 (  2.08) &  -1.25 (  2.37) &  -1.29 (  1.39) \\ 
  1363 &  53053 &    104 &   0.18 &   0.27 &   1.12 (  0.38) &   55.6 &   \dots (\dots) &    0.0 (  1.55) &   -0.05 (  9.05) &   0.17 (  7.07) &   0.21 (\dots) \\ 
  1373 &  53063 &    255 &   0.03 &   0.13 &   \dots (\dots) &    0.0 &   0.0 (  2.30)     &    \dots (\dots) &    \dots (\dots) &   \dots (\dots) &   \dots (\dots) \\ 
  1390 &  53142 &    385 &   0.13 &   0.29 &   2.08 (  0.49) &   26.1 &   \dots (\dots) &    \dots (\dots) &    0.07 (  2.41) &   0.05 (  1.68) &   0.07 (\dots) \\ 
  1416 &  52875 &    490 &   0.07 &   0.19 &   2.18 (  1.17) &   14.4 &   \dots (\dots) &   -0.03 (  1.77) &    \dots (\dots) &   \dots (\dots) &  0.0 (  2.92) \\ 
  1423 &  53167 &    528 &   0.20 &   0.33 &   1.73 (  0.50) &   42.2 &   \dots (\dots) &    \dots (\dots) &   -0.54 (  1.35) &  -0.18 (  1.77) &  -0.30 (\dots) \\ 
  1429 &  52990 &    336 &   0.07 &   0.15 &   1.25 (  0.16) &   42.7 &   0.0 (  2.61)     &    0.0 (  2.78) &    0.67 (  1.63) &   0.06 (  2.11) &   1.54 (  9.57) \\ 
  1604 &  53078 &    459 &   0.08 &   0.17 &   0.82 (  0.43) &   42.2 &   \dots (\dots) &    0.94 (  6.10) &    0.49 (  1.32) &   1.61 (  1.98) &   \dots (\dots) \\ 
  1607 &  53083 &     48 &   0.08 &   0.15 &   \dots (\dots) &    0.0 &   \dots (\dots)  &   -0.63 (  6.30) &    \dots (\dots) &   \dots (\dots) &   0.0 (  1.39) \\ 
  1618 &  53116 &    285 &   0.04 &   0.17 &   \dots (\dots) &    0.0 &   \dots (\dots) &    \dots (\dots) &    \dots (\dots) &   \dots (\dots) &   0.0 (  2.32) \\ 
  1620 &  53137 &    175 &   0.07 &   0.10 &   1.00 (  0.21) &   47.7 &   0.0 (  8.61)     &   -0.23 (  1.69) &    0.05 (  5.04) &  -0.17 ( 23.12) &  -0.12 (\dots) \\ 
  1677 &  53148 &    262 &   0.19 &   0.31 &   \dots (\dots) &    0.0 &   \dots (\dots) &    \dots (\dots) &    \dots (\dots) &   \dots (\dots) &   \dots (\dots) \\ 
  1746 &  53062 &    533 &   0.14 &   0.25 &   1.21 (  0.38) &   50.1 &   \dots (\dots) &    0.39 (  4.83) &    0.80 (  1.64) &   0.36 (  1.92) &  -0.23 (\dots) \\ 
\enddata
\end{deluxetable}

\begin{deluxetable}{l l l l l l l l l l l}
\tabletypesize{\footnotesize}
\tablewidth{0pt.}
\tablecaption{ \label{t:stack} The optical depths of the stacked SDSS filters. These values have not been corrected for inclination.}
\tablehead{ \colhead{plate}  &\colhead{mjd} & \colhead{fiber} & \colhead{$z_{fg}$} & \colhead{$z_{bg}$} & \colhead{R} & \colhead{$\tau_{gr}$} &\colhead{$\tau_{ri}$} &\colhead{$\tau_{iz}$} &\colhead{$\tau_{gri}$} &\colhead{$\tau_{griz}$} }
\startdata
   267 &  51608 &    499 &   0.04 &   0.09 &   \dots (\dots) &   \dots (\dots) &    \dots (\dots) &    \dots (\dots) &   \dots (\dots) &  -0.13 (4.83) \\ 
   273 &  51957 &    323 &   0.11 &   0.18 &   \dots (\dots) &   \dots (\dots) &    \dots (\dots) &    \dots (\dots) &   \dots (\dots) &   \dots (\dots) \\ 
   275 &  51910 &    233 &   0.03 &   0.09 &   0.80 (0.09) &   \dots (\dots) &    0.30 (18.64) &    0.05 (7.67) &   0.31 (5.25) &   0.05 (16.51) \\ 
   281 &  51614 &    624 &   0.05 &   0.20 &   \dots (\dots) &   \dots (\dots) &    \dots (\dots) &    \dots (\dots) &   \dots (\dots) &   \dots (\dots) \\ 
   282 &  51658 &    120 &   0.07 &   0.26 &   1.52 (0.24) &   \dots (\dots) &   -0.03 (1.43) &   -0.00 (1.60) &   \dots (\dots) &   0.00 (15.92) \\ 
   285 &  51930 &     15 &   0.02 &   0.14 &   \dots (\dots) &   \dots (\dots) &    \dots (\dots) &    \dots (\dots) &   \dots (\dots) &   \dots (\dots) \\ 
   313 &  51673 &    546 &   0.04 &   0.30 &   \dots (\dots) &   \dots (\dots) &    \dots (\dots) &    \dots (\dots) &   \dots (\dots) &   \dots (\dots) \\ 
   367 &  51997 &    222 &   0.08 &   0.14 &   \dots (\dots) &   \dots (\dots) &    \dots (\dots) &    \dots (\dots) &   \dots (\dots) &   \dots (\dots) \\ 
   403 &  51871 &    586 &   0.14 &   0.17 &   \dots (\dots) &   \dots (\dots) &    \dots (\dots) &    \dots (\dots) &   \dots (\dots) &   \dots (\dots) \\ 
   408 &  51821 &    336 &   0.07 &   0.12 &   \dots (\dots) &   \dots (\dots) &    0.39 (1.45) &    \dots (\dots) &   \dots (\dots) &   \dots (\dots) \\ 
   435 &  51882 &    118 &   0.23 &   0.26 &   \dots (\dots) &   \dots (\dots) &    \dots (\dots) &    \dots (\dots) &   \dots (\dots) &   \dots (\dots) \\ 
   436 &  51883 &    638 &   0.04 &   0.36 &   1.00 (0.18) &   \dots (\dots) &    0.15 (15.17) &    0.19 (6.96) &   \dots (\dots) &   0.29 (8.28) \\ 
   440 &  51885 &    370 &   0.18 &   0.25 &   \dots (\dots) &   \dots (\dots) &    \dots (\dots) &    \dots (\dots) &   \dots (\dots) &   \dots (\dots) \\ 
   464 &  51908 &    211 &   0.12 &   0.20 &   \dots (\dots) &   \dots (\dots) &    \dots (\dots) &    \dots (\dots) &   \dots (\dots) &   \dots (\dots) \\ 
   474 &  52000 &    548 &   0.21 &   0.30 &   \dots (\dots) &   \dots (\dots) &    \dots (\dots) &    \dots (\dots) &   \dots (\dots) &   \dots (\dots) \\ 
   484 &  51907 &    616 &   0.16 &   0.25 &   \dots (\dots) &   \dots (\dots) &    \dots (\dots) &    \dots (\dots) &   \dots (\dots) &   \dots (\dots) \\ 
   487 &  51943 &    305 &   0.12 &   0.25 &   3.51 (0.37) &   \dots (\dots) &   -0.24 (1.80) &   -0.09 (4.14) &   \dots (\dots) &  -0.09 (3.63) \\ 
   497 &  51989 &     73 &   0.20 &   0.40 &   1.27 (0.36) &   \dots (\dots) &    0.04 (1.79) &    0.10 (2.47) &   \dots (\dots) &   \dots (\dots) \\ 
   499 &  51988 &    152 &   0.11 &   0.29 &   \dots (\dots) &   \dots (\dots) &   -0.42 (1.65) &    \dots (\dots) &   \dots (\dots) &   \dots (\dots) \\ 
   511 &  52636 &    313 &   0.13 &   0.27 &   \dots (\dots) &   \dots (\dots) &    0.52 (2.55) &    \dots (\dots) &   0.21 (1.04) &   \dots (\dots) \\ 
   512 &  51992 &    302 &   0.05 &   0.23 &   0.98 (0.24) &   \dots (\dots) &    0.00 (18.95) &   -0.06 (26.34) &   \dots (\dots) &   0.01 (17.92) \\ 
   529 &  52025 &    408 &   0.02 &   0.17 &   \dots (\dots) &   \dots (\dots) &    \dots (\dots) &    \dots (\dots) &   \dots (\dots) &   \dots (\dots) \\ 
   550 &  51959 &    624 &   0.17 &   0.20 &   \dots (\dots) &   \dots (\dots) &    \dots (\dots) &    \dots (\dots) &   \dots (\dots) &   \dots (\dots) \\ 
   628 &  52083 &    262 &   0.03 &   0.12 &   \dots (\dots) &   \dots (\dots) &    \dots (\dots) &    \dots (\dots) &   \dots (\dots) &   \dots (\dots) \\ 
   629 &  52051 &     26 &   0.12 &   0.35 &   0.76 (0.30) &   \dots (\dots) &    0.19 (1.66) &    0.37 (11.43) &   \dots (\dots) &   \dots (\dots) \\ 
   637 &  52174 &    334 &   0.09 &   0.18 &   1.12 (0.50) &   \dots (\dots) &    0.00 (5.20) &    0.22 (2.16) &   \dots (\dots) &   \dots (\dots) \\ 
   671 &  52206 &    347 &   0.04 &   0.28 &   3.88 (0.14) &   \dots (\dots) &    0.10 (2.09) &   -0.33 (2.66) &   \dots (\dots) &  -0.84 (2.07) \\ 
   765 &  52254 &    179 &   0.04 &   0.21 &   \dots (\dots) &   \dots (\dots) &    \dots (\dots) &    \dots (\dots) &   \dots (\dots) &   \dots (\dots) \\ 
   782 &  52320 &    306 &   0.15 &   0.21 &   \dots (\dots) &   \dots (\dots) &    \dots (\dots) &    \dots (\dots) &   \dots (\dots) &   \dots (\dots) \\ 
   790 &  52441 &    527 &   0.03 &   0.14 &   \dots (\dots) &   \dots (\dots) &   -0.03 (1.43) &    \dots (\dots) &   \dots (\dots) &   \dots (\dots) \\ 
   792 &  52353 &    432 &   0.14 &   0.20 &   1.88 (0.45) &   \dots (\dots) &    0.15 (2.90) &    0.07 (1.81) &   \dots (\dots) &   0.08 (2.16) \\ 
   808 &  52556 &     39 &   0.21 &   0.37 &   \dots (\dots) &   \dots (\dots) &    \dots (\dots) &    \dots (\dots) &   \dots (\dots) &   \dots (\dots) \\ 
   845 &  52381 &    270 &   0.07 &   0.32 &   \dots (\dots) &   \dots (\dots) &    \dots (\dots) &    \dots (\dots) &   \dots (\dots) &   \dots (\dots) \\ 
   848 &  52669 &    602 &   0.05 &   0.22 &   1.44 (0.16) &   \dots (\dots) &    0.22 (1.84) &    0.65 (1.67) &   \dots (\dots) &   0.30 (1.71) \\ 
   860 &  52319 &    451 &   0.04 &   0.16 &   1.54 (0.22) &   0.35 (1.93) &    0.28 (1.73) &    0.37 (1.69) &   0.37 (1.29) &   0.42 (1.93) \\ 
   864 &  52320 &     91 &   0.04 &   0.26 &   1.89 (0.25) &   \dots (\dots) &   -0.17 (3.73) &   -0.20 (3.50) &   \dots (\dots) &  -0.05 (3.88) \\ 
   877 &  52353 &    458 &   0.04 &   0.06 &   \dots (\dots) &   \dots (\dots) &    \dots (\dots) &    \dots (\dots) &   \dots (\dots) &   \dots (\dots) \\ 
   882 &  52370 &    122 &   0.06 &   0.18 &   \dots (\dots) &   \dots (\dots) &    \dots (\dots) &    \dots (\dots) &   \dots (\dots) &   \dots (\dots) \\ 
   883 &  52430 &    366 &   0.05 &   0.13 &   2.78 (0.55) &   \dots (\dots) &   -0.18 (17.96) &   -0.09 (2.66) &   \dots (\dots) &   \dots (\dots) \\ 
   884 &  52374 &    242 &   0.04 &   0.17 &   \dots (\dots) &   \dots (\dots) &    \dots (\dots) &    \dots (\dots) &   \dots (\dots) &   \dots (\dots) \\ 
   889 &  52663 &    408 &   0.06 &   0.16 &   \dots (\dots) &   \dots (\dots) &    \dots (\dots) &    \dots (\dots) &   \dots (\dots) &   \dots (\dots) \\ 
   896 &  52592 &    463 &   0.03 &   0.11 &   2.10 (0.20) &   0.35 (2.16) &    0.16 (6.66) &    0.02 (2.51) &   0.47 (1.58) &   0.09 (4.17) \\ 
   913 &  52433 &    151 &   0.08 &   0.16 &   \dots (\dots) &   \dots (\dots) &   -0.10 (1.79) &    \dots (\dots) &   \dots (\dots) &   \dots (\dots) \\ 
   931 &  52619 &     84 &   0.18 &   0.28 &   \dots (\dots) &   \dots (\dots) &    \dots (\dots) &    \dots (\dots) &   \dots (\dots) &   \dots (\dots) \\ 
   958 &  52410 &    194 &   0.15 &   0.31 &   4.08 (0.77) &   \dots (\dots) &   -0.55 (1.84) &    0.04 (2.64) &   \dots (\dots) &   0.00 (1.03) \\ 
   970 &  52413 &    408 &   0.30 &   0.38 &   \dots (\dots) &   \dots (\dots) &    \dots (\dots) &    \dots (\dots) &   \dots (\dots) &   \dots (\dots) \\ 
   980 &  52431 &    300 &   0.13 &   0.29 &   0.99 (0.28) &   \dots (\dots) &    0.00 (1.73) &   -0.27 (115.35) &   \dots (\dots) &  -0.28 (11.09) \\ 
  1000 &  52643 &    337 &   0.09 &   0.32 &   \dots (\dots) &   \dots (\dots) &    \dots (\dots) &    \dots (\dots) &   \dots (\dots) &   \dots (\dots) \\ 
  1006 &  52708 &    624 &   0.05 &   0.13 &   1.94 (0.25) &   \dots (\dots) &    0.39 (1.83) &    0.21 (1.92) &   \dots (\dots) &   0.40 (38.38) \\ 
  1007 &  52706 &    519 &   0.05 &   0.20 &   \dots (\dots) &   \dots (\dots) &    \dots (\dots) &    \dots (\dots) &   \dots (\dots) &   \dots (\dots) \\ 
  1160 &  52674 &    468 &   0.24 &   0.29 &   \dots (\dots) &   \dots (\dots) &    \dots (\dots) &    \dots (\dots) &   \dots (\dots) &   \dots (\dots) \\ 
  1215 &  52725 &    288 &   0.27 &   0.33 &   \dots (\dots) &   \dots (\dots) &    \dots (\dots) &    \dots (\dots) &   \dots (\dots) &   \dots (\dots) \\ 
  1230 &  52672 &    639 &   0.05 &   0.16 &   1.61 (0.48) &   \dots (\dots) &    0.09 (13.43) &    0.08 (1.68) &   \dots (\dots) &   0.10 (1.29) \\ 
  1235 &  52734 &    164 &   0.10 &   0.33 &   2.89 (0.71) &   \dots (\dots) &    0.03 (1.11) &   -0.09 (2.04) &   \dots (\dots) &  -0.07 (3.27) \\ 
  1269 &  52937 &    485 &   0.03 &   0.19 &   \dots (\dots) &   \dots (\dots) &    \dots (\dots) &    \dots (\dots) &   \dots (\dots) &   \dots (\dots) \\ 
  1280 &  52738 &    125 &   0.08 &   0.21 &   \dots (\dots) &   \dots (\dots) &    \dots (\dots) &    \dots (\dots) &   \dots (\dots) &   \dots (\dots) \\ 
  1282 &  52759 &    630 &   0.12 &   0.28 &   0.52 (0.29) &   0.32 (4.08) &    0.25 (2.51) &    0.09 (3.18) &   \dots (\dots) &   0.12 (1.75) \\ 
  1310 &  53033 &    202 &   0.11 &   0.18 &   \dots (\dots) &   \dots (\dots) &    \dots (\dots) &    \dots (\dots) &   \dots (\dots) &   \dots (\dots) \\ 
  1317 &  52765 &      9 &   0.10 &   0.20 &   \dots (\dots) &   \dots (\dots) &    \dots (\dots) &    \dots (\dots) &   \dots (\dots) &   \dots (\dots) \\ 
  1321 &  52764 &    461 &   0.15 &   0.22 &   4.61 (0.40) &   \dots (\dots) &    0.16 (1.95) &    0.09 (6.27) &   \dots (\dots) &   0.00 (1.99) \\ 
  1324 &  53088 &     50 &   0.04 &   0.23 &   1.29 (0.21) &   \dots (\dots) &    0.04 (129.61) &    0.02 (11.83) &   \dots (\dots) &   0.00 (7.32) \\ 
  1325 &  52762 &    313 &   0.12 &   0.21 &   \dots (\dots) &   \dots (\dots) &    \dots (\dots) &    \dots (\dots) &   \dots (\dots) &   \dots (\dots) \\ 
  1327 &  52781 &    252 &   0.08 &   0.14 &   \dots (\dots) &   \dots (\dots) &    \dots (\dots) &    \dots (\dots) &   \dots (\dots) &   \dots (\dots) \\ 
  1332 &  52781 &    469 &   0.28 &   0.32 &   3.87 (0.90) &   \dots (\dots) &   -0.99 (3.02) &   -1.28 (21.59) &   \dots (\dots) &  -0.99 (1.44) \\ 
  1363 &  53053 &    104 &   0.18 &   0.27 &   2.02 (0.44) &   \dots (\dots) &    0.25 (11.95) &    0.21 (1.32) &   \dots (\dots) &   \dots (\dots) \\ 
  1373 &  53063 &    255 &   0.03 &   0.13 &   \dots (\dots) &   \dots (\dots) &    \dots (\dots) &    \dots (\dots) &   \dots (\dots) &   \dots (\dots) \\ 
  1390 &  53142 &    385 &   0.13 &   0.29 &   1.17 (0.54) &   \dots (\dots) &    0.17 (4.26) &    0.23 (4.71) &   \dots (\dots) &   \dots (\dots) \\ 
  1416 &  52875 &    490 &   0.07 &   0.19 &   \dots (\dots) &   \dots (\dots) &    \dots (\dots) &    \dots (\dots) &   \dots (\dots) &   \dots (\dots) \\ 
  1423 &  53167 &    528 &   0.20 &   0.33 &   \dots (\dots) &   \dots (\dots) &   -0.31 (1.17) &    \dots (\dots) &   \dots (\dots) &   \dots (\dots) \\ 
  1429 &  52990 &    336 &   0.07 &   0.15 &   1.27 (0.15) &   0.00 (4.62) &    0.48 (2.55) &   -1.11 (8.44) &   0.97 (3.81) &   0.78 (1.85) \\ 
  1604 &  53078 &    459 &   0.08 &   0.17 &   0.90 (0.52) &   \dots (\dots) &    0.64 (8.19) &   -0.04 (4.06) &   \dots (\dots) &   0.82 (1.06) \\ 
  1607 &  53083 &     48 &   0.08 &   0.15 &   \dots (\dots) &   \dots (\dots) &    \dots (\dots) &    \dots (\dots) &   \dots (\dots) &   \dots (\dots) \\ 
  1618 &  53116 &    285 &   0.04 &   0.17 &   \dots (\dots) &   \dots (\dots) &    \dots (\dots) &    \dots (\dots) &   \dots (\dots) &   \dots (\dots) \\ 
  1620 &  53137 &    175 &   0.07 &   0.10 &   1.33 (0.29) &   \dots (\dots) &   -0.11 (3.35) &   -0.11 (1.32) &   \dots (\dots) &  -0.07 (11.45) \\ 
  1677 &  53148 &    262 &   0.19 &   0.31 &   \dots (\dots) &   0.44 (2.03) &    \dots (\dots) &    \dots (\dots) &   \dots (\dots) &   \dots (\dots) \\ 
  1746 &  53062 &    533 &   0.14 &   0.25 &   1.14 (0.35) &   \dots (\dots) &    0.68 (8.73) &    0.44 (2.08) &   \dots (\dots) &   \dots (\dots) 
\enddata
\end{deluxetable}

\begin{deluxetable}{l l l l l l l l l l l l}
\tablewidth{0pt.}
\tablecaption{ \label{t:opt_exp} Exponential fit values to optimum aperture optical depths}
\tablehead{ \colhead{filter} & \colhead{$0.01 < z < 0.1$} &  & & \colhead{$0.1 < z < 0.2$} &  & \\
 & \colhead{$\tau_0$} & \colhead{h} &\colhead{rms} & \colhead{$\tau_0$} & \colhead{h} & \colhead{rms} \\
}
\startdata
u 	& 0.1 & -1.8	& 0   		& \dots	& \dots	& \dots \\
g	& 1.4 & 3.9   	& 1.0       	& 0.3		& -16.7	& 0.6 \\
i	& 0.3 & -70.3    	& 1.2		& 0.2		& -3.4	& 0.9 \\
r	& 0.7 & 7.6    	& 1.2		& 0.3		& -7.2	& 0.8 \\
z	& 0.4 & -648.5	& 1.1		& 0.7		& 6.6		& 1.0 \\
\enddata
\end{deluxetable}

\begin{deluxetable}{l l l l l l l l l l l l}
\tablewidth{0pt.}
\tablecaption{ \label{t:vis_exp} Exponential fit values to visual aperture optical depths}
\tablehead{ \colhead{filter} & \colhead{$0.01 < z < 0.1$} &  & & \colhead{$0.1 < z < 0.2$} &  & \\
 & \colhead{$\tau_0$} & \colhead{h} &\colhead{rms} & \colhead{$\tau_0$} & \colhead{h} & \colhead{rms} \\
}
\startdata
u 	& \dots	& \dots	& \dots 	& \dots	& \dots	& \dots \\
g 	& 0.04	& -1.0	& 1.0		& 0.4 	& 17.8	& 0.6\\
i 	& 0.1		& -2.1	& 1.2       	& 0.3 	& 6.2		& 0.9\\
r 	& 0.1		& -3.8	& 1.2       	& 2.1		& 0.5		& 0.8\\
z	& 0.3		& -5.6	& 1.1       	& 0.4		& 9.3		& 1.0\\
\enddata
\end{deluxetable}

\begin{deluxetable}{l l l l}
\tablewidth{0pt.}
\tablecaption{ \label{t:ri_exp} Exponential fit values to visual aperture optical depths in the r+i }
\tablehead{ \colhead{Redshift } & \colhead{$\tau_0$} & \colhead{h} &\colhead{rms} \\}
\startdata
$z<0.01$		& 0.9		& 1.3		&  0.9 \\
(arm) &	0.8	& 2.3		& 0.9 \\
(disk) &	1.2	& 0.8		& 0.7 \\
$0.01<z<0.1$	& 0.2		& 2.9		& 1.2 \\
$0.1<z<0.2$	& 0.3 	& 7.4		& 0.7 \\
\enddata
\end{deluxetable}

\end{document}